\documentclass{article}

     \usepackage[nonatbib,preprint]{neurips_2019}

\usepackage[utf8]{inputenc} 
\usepackage[T1]{fontenc}    
\usepackage{hyperref}       
\usepackage{url}            
\usepackage{booktabs}       
\usepackage{amsfonts}       
\usepackage{nicefrac}       
\usepackage{microtype}      

\usepackage[linesnumbered,ruled,vlined]{algorithm2e}
\usepackage{float}
\usepackage{amsbsy}
\usepackage{amssymb}
\usepackage{amsmath}
\usepackage{amsthm}
\usepackage[]{graphicx}
\usepackage{setspace}
\usepackage{esint}
\usepackage{comment}
\usepackage{color}
\usepackage{url}
\usepackage{hyperref}
\usepackage[capitalize]{cleveref}  
\usepackage{subcaption}

\usepackage[suppress]{color-edits}
\addauthor{vs}{blue}
\addauthor{mo}{red}
\addauthor{md}{red}

\newcommand{\kibitz}[2]{\ifnum\Comments=1{\color{#1}{#2}}\fi}

\newcommand{\E}{\mathbb{E}}

\DeclareMathOperator*{\arginf}{arg\,inf}

\newcommand{\R}{\mathbb{R}}

\newcommand{\D}{\mathcal D}

\renewcommand{\Pr}{\ensuremath{\mathrm{Pr}}}

\newcommand{\ba}{\begin{array}}
\newcommand{\ea}{\end{array}}
\newcommand{\bs}{\begin{align}\begin{split}\nonumber}
\newcommand{\bsnumber}{\begin{align}\begin{split}}
\newcommand{\es}{\end{split}\end{align}}

\newtheorem{assumption}{ASSUMPTION}

\newcommand{\ldot}[2]{\langle #1, #2 \rangle}

\newcommand{\Var}{\ensuremath{\text{Var}}}

\def\balign#1\ealign{\begin{align}#1\end{align}}
\def\balignat#1\ealign{\begin{alignat}#1\end{alignat}}
\def\bitemize#1\eitemize{\begin{itemize}#1\end{itemize}}
\def\benumerate#1\eenumerate{\begin{enumerate}#1\end{enumerate}}
\newenvironment{talign}
 {\csname align\endcsname}
 {\endalign}
\def\balignt#1\ealignt{\begin{talign}#1\end{talign}}%
\newtheorem{theorem}{Theorem}
\newtheorem{corollary}[theorem]{Corollary}
\newtheorem{lemma}[theorem]{Lemma}
\newtheorem{definition}{Definition}

\newcommand\blfootnote[1]{%
  \begingroup
  \renewcommand\thefootnote{}\footnote{#1}%
  \addtocounter{footnote}{-1}%
  \endgroup
}

\usepackage{hyperref}
\renewcommand{\textstyle}[1]{#1}

\title{Machine Learning Estimation of Heterogeneous Treatment Effects with Instruments}

\author{%
Vasilis Syrgkanis\\
Microsoft Research\\
\texttt{vasy@microsoft.com}
\And
Victor Lei\\
TripAdvisor\\
\texttt{vlei@tripadvisor.com}
\And
Miruna Oprescu\\
Microsoft Research\\
\texttt{moprescu@microsoft.com}
\And
Maggie Hei\\
Microsoft Research\\
\texttt{Maggie.Hei@microsoft.com}
\And
Keith Battocchi\\
Microsoft Research\\
\texttt{kebatt@microsoft.com}
\And
Greg Lewis\\
Microsoft Research\\
\texttt{glewis@microsoft.com}
}

\begin{document}

\maketitle

\begin{abstract}
    We consider the estimation of heterogeneous treatment effects with arbitrary machine learning methods in the presence of unobserved confounders with the aid of a valid instrument. Such settings arise in A/B tests with an intent-to-treat structure, where the experimenter randomizes over which user will receive a recommendation to take an action, and we are interested in the effect of the downstream action. We develop a statistical learning approach to the estimation of heterogeneous effects, reducing the problem to the minimization of an appropriate loss function that depends on a set of auxiliary models (each corresponding to a separate prediction task). The reduction enables the use of all recent algorithmic advances (e.g. neural nets, forests). We show that the estimated effect model is robust to estimation errors in the auxiliary models, by showing that the loss satisfies a Neyman orthogonality criterion. Our approach can be used to estimate projections of the true effect model on simpler hypothesis spaces. When these spaces are parametric, then the parameter estimates are asymptotically normal, which enables construction of confidence sets. We applied our method to estimate the effect of membership on downstream webpage engagement on TripAdvisor, using as an instrument an intent-to-treat A/B test among 4 million TripAdvisor users, where some users received an easier membership sign-up process. We also validate our method on synthetic data and on public datasets for the effects of schooling on income.\blfootnote{Prototype code for all the algorithms presented and the synthetic data experimental study can be found at \href{https://github.com/microsoft/EconML/tree/master/prototypes/dml_iv}{https://github.com/Microsoft/EconML/tree/master/prototypes/dml\_iv}}
\end{abstract}

\section{Introduction}


A/B testing is the gold standard of causal inference. But even when A/B testing is feasible, estimating the effect of a treatment on an outcome might not be a straightforward task. One major difficulty is non-compliance: even if we randomize what treatment to recommend to a subject, the subject might not comply with the recommendation due to unobserved factors and follow the alternate action. \moedit{The impact that unobserved factors might have on the measured outcome} is a source of endogeneity and can lead to biased estimates of the effect. 
\moedit{This problem arises in large scale} data problems in the digital economy; when optimizing a digital service, we might often want to estimate the effect of some action taken by our users on downstream metrics. However, the service cannot force users to comply, but can only find means of incentivizing or recommending the action. The unobserved factors of compliance can lead to biased estimates if we consider the takers and not takers as exogenously assigned and employ machine learning approaches to estimate the potentially heterogeneous effect of the action on the downstream metric.

The problem can be solved by using the technique of instrumental variable (IV) regression: as long as the recommendation increases the probability of taking the treatment, then we know that there is at least some fraction of users that were assigned the treatment ``exogeneously''. IV regression parses out this population of ``exogenously treated'' users and estimates an effect based solely on them.

Most classical IV approaches estimate a constant average treatment effect. However, to make personalized policy decisions (an emerging trend in most digital services) one might want to estimate a heterogeneous effect based on observable characteristics of the user. The latter is a daunting task, as we seek to estimate a function of observable characteristics as opposed to a single number. Hence, statistical power is at stake. Even estimating an ATE is non-trivial when effect and compliance are correlated through observables. The emergence of large data-sets in the digital economy alleviates this concern; with A/B tests running on millions of users it is possible to estimate complex heterogeneous effect models, even if compliance levels are relatively weak. Moreover, as we control for more and more observable features of the user, we also reduce the risk that correlation between effect and compliance is stemming from unobserved factors.

This leads to the question this work seeks to answer: how can we blend the power of modern machine learning approaches (e.g. random forests, gradient boosting, penalized regressions, neural networks) with instrumental variable methods, so as to estimate complex heterogeneous effect rules. Recent work at the intersection of machine learning and econometrics has proposed powerful methods for estimating the effect of a treatment on an outcome, while using machine learning methods for learning nuisance models that help de-bias the final effect rule. However, the majority of the work has either focused on 1) estimating average treatment effects or low dimensional parametric effect models (e.g. the double machine learning approach of \cite{Chernozhukov2018}), 2) developing new algorithms for estimating fully non-parametric models of the effect (e.g. the IV forest method of \cite{athey2019generalized}, the DeepIV method of \cite{pmlr-v70-hartford17a}), 3) assuming that the treatment is exogenous once we condition on the observable features and reducing the problem to an appropriate square loss minimization framework (see e.g. \cite{nie2017quasi,kunzel2017meta}). 

Nevertheless, a general reduction of IV based machine learning estimation of heterogeneous effects to a more standard statistical learning problem that can incorporate existing algorithms in a black-box manner has not been formulated in prior work. In fact, the work of \cite{nie2017quasi} leaves this as an open question. Such a reduction can help us leverage the recent algorithmic advances in statistical learning theory so as to work with large data-sets. Our work proposes the reduction of heterogeneous effects estimation via instruments to a square loss minimization problem over a hypothesis space. This enables us to learn not only the true heterogeneous effect model, but also the projections of the true model in simpler hypothesis spaces for interpretability. Moreover, our work leverages recent advances in statistical learning with nuisance functions \cite{chernozhukov2018plug,foster2019orthogonal}, to show that the mean squared error (MSE) of the learned model is robust to the estimation error of auxiliary models that need to be estimated (as is standard in IV regression). Thus we achieve MSE rates where the leading term depends only on the sample complexity of the hypothesis space of the heterogeneous effect model. 

Some advantages of reducing our problem to a set of standard regression problems include being able to use existing algorithms and implementations, as well as recent advances of interpretability in machine learning. For instance, in our application we deploy the SHAP framework \cite{SHAP1,SHAP2} to interpret random forest based models of the heterogeneous effect.
Furthermore, when the hypothesis space is low dimensional and parametric then our approach falls in the setting studied by prior work of \cite{Chernozhukov2018} and, hence, not only MSE rates but also confidence interval construction is relatively straightforward. This enables hypothesis testing on parametric projections of the true effect model.

We apply our approach to an intent-to-treat A/B test among 4 million users on a major travel webpage so as to estimate the effect of membership on downstream engagement. We identify sources of heterogeneity that have policy implications on which users the platform should engage more and potentially how to re-design the recommendation to target users with large effects. We validate the findings on a different cohort in a separate experiment among 10 million users on the same platform. Even though the new experiment was deployed on a much broader and different cohort, we identify common leading factors of effect heterogeneity, hence confirming our findings. As a robustness check we create semi-synthetic data with similar features and marginal distributions of variables as the real data, but where we know the ground truth. We find that our method performs well both in terms of MSE, identifying the relevant factors and coverage of the confidence intervals. 

Finally, we apply our method to a more traditional IV application: estimating the effect of schooling on wages. We use a well studied public data set and observe that our approach automatically identifies sources of heterogeneity that were previously uncovered using more structural approaches. We also validate our method in this application on semi-synthetic data that emulate the true data.

\section{Estimation of Heterogeneous Treatment Effects with Instruments}

We consider estimation of heterogeneous treatment effects with respect to a set of features $X$, of an endogenous treatment $T$ on an outcome $Y$ with an instrument $Z$. For simplicity of exposition, we will restrict attention to the case where $Y, Z$ and $T$ are scalar variables, but several of our results extend to the case of multi-dimensional treatments and instruments. $Z$ is an instrumental variable if it has an effect on the treatment but does not have a direct effect on the outcome other than through the treatment. More formally, we assume the following moment condition:
\begin{equation}
    \E[Y - \theta_0(X) T - f_0(X) \mid Z, X]=0
\end{equation}
Equivalently we assume that: $Y = \theta_0(X)\, T + f_0(X) + e$, with  $\E[e \mid Z, X] = 0$.
We allow for the presence of confounders, i.e. $e$ could be correlated with $T$ via some unobserved common factor $\nu$. However, our exclusion restriction on the instrument implies that the residual is mean zero conditional on the instrument. Together with the fact that the instrument also has an effect on the treatment at any value of the feature $X$, i.e.:
$\Var(\E[T\mid Z, X] \mid X) \geq \lambda$,
allows us to identify the heterogeneous effect function $\theta_0(X)$. We focus on the case where the effect is linear in the treatment $T$, which is wlog in the binary treatment setting, which is our main application, and since our goal is to focus on the non-linearity wrt $X$ (this greatly simplifies our problem, see \cite{Chen2015,ChenPouzo,NeweyPowell,hartford2017deep}).

Given $n$ i.i.d. samples from the data generating process, our goal is to estimate a model $\hat{\theta}(X)$ that achieves small expected mean-squared-error, i.e.: $\E[\|\hat{\theta}-\theta_0\|_2] := \E[(\hat{\theta}(X) - \theta_0(X))^2] \leq R_{n}$.
Since the true $\theta_0$ function can be very complex and \moedit{difficult} to estimate in finite samples, we \moedit{are also} interested in estimating projections of the true $\theta_0$ on simpler hypothesis spaces $\Theta_{\pi}$. \moedit{Projections are also useful for interpretability}: one might want to understand what is the best linear projection of $\theta_0(X)$ on $X$, i.e. $\alpha_0 = \arg\min_{\alpha}  \E[(\ldot{\alpha}{X}- \theta_0(X))^2]$. In this case we will denote with $\theta_*$ the projection of $\theta_0$ on $\Theta_{\pi}$, i.e. $\theta_*=\arg\min_{\theta\in \Theta_{\pi}}  \E[(\theta(X) - \theta_0(X))^2]$ and our goal would be to achieve small mean squared error with respect to $\theta_*$. When $\theta_*$ is a low dimensional parametric class (e.g. a linear function on a low-dimensional feature space or a constant function), we are also interested in performing inference; i.e. constructing confidence intervals that asymptotically contain the correct parameter with probability equal to some target confidence level.

\vsdelete{Our primary technical development is to formulate a loss minimization based approach that enables estimation of $\hat{\theta}(X)$ at MSE rates with respect to $\theta_*$, that are primarily driven by the sample complexity of the function space $\Theta_{\pi}$. The latter is not trivial, since estimating $\hat{\theta}_*$ essentially requires first estimating $\theta_0$, which depends on the sample complexity of the true function space. Moreover, estimating $\theta_0$ also depends on estimating $f_0$ and $h_0(Z,X)=\E[T\mid Z, X]$. These functions could be in turn much more complex than $\Theta_{\pi}$. To achieve the robustness property, we will use the recent framework of orthogonal statistical learning of \cite{foster2019orthogonal}, and formulate a loss function for learning an estimate of $\theta_*$, that is orthogonal with respect to the nuisance functions that also need to be estimated from the data. Moreover, we want our loss function to be strongly convex with respect to $\theta(X)$ so that fast estimation rates can be achieved. Then we can invoke the results of \cite{foster2019orthogonal} to get fast convergence rates for several function classes of interest.}

\subsection{Warm-Up: Estimating the Average Treatment Effect (ATE)}
\label{sec:warm} For estimation of the average treatment effect (ATE), \emph{assuming that either there is no effect heterogeneity with respect to $X$ or there is no heterogeneous compliance with respect to $X$}, \cite{Chernozhukov2018} propose a method for estimating the ATE that \moedit{solves} the empirical analogue of the following moment equation:
\begin{equation}
    \E[ (Y - \E[Y\mid X] - \theta (T - \E[T\mid X]))\, (Z - \E[Z\mid X])] =0
\end{equation}
This moment function is orthogonal to all the functions $q_0(X) = \E[Y \mid X]$, $p_0(X)=\E[T\mid X]$ and $r_0(X)=\E[Z\mid X]$ that also need to be estimated from data. This moment avoids the estimation of the expected $T$ conditional on $Z, X$ and satisfies an orthogonality condition that enables robustness of the estimate $\hat{\theta}=\frac{\E_n[(Y - \hat{q}(X))\, (Z - \hat{r}(X))]}{\E_n[(T - \hat{p}(X))\, (Z - \hat{r}(X))]}$,
to \moedit{errors in the nuisance} estimates $\hat{q}, \hat{r}$ and $\hat{p}$. \moedit{The} estimate is asymptotically normal with variance equal to the variance of the method if the estimates were the correct ones, assuming that the mean squared error of these estimates decays at least at a rate of $n^{-1/4}$ (see \cite{Chernozhukov2018} for more details). This result requires that the nuisance estimates are fitted in a \emph{cross-fitting} manner, i.e. we use half of the data to fit a model for each of these functions and then predict the values of the model on the other half of the samples. We refer to this algorithm as DMLATEIV.\footnote{For Double Machine Learning ATE estimation with Instrumental Variables.} \vsdelete{\moedit{DMLATEIV} is algorithmically equivalent to a two stage algorithm, where we run a linear regression to predict $T-\hat{p}(X)$ from $Z-\hat{r}(X)$ and then running a linear regression between $Y-\hat{q}(X)$ and the predicted $\E[T-\hat{p}(X)\mid Z-\hat{r}(X)]$. The coefficient in the final regression is exactly equal to the estimate $\hat{\theta}$ (by simple algebraic manipulations).}

\textbf{Inconsistency under Effect and Compliance Heterogeneity}
The above estimate $\hat{\theta}$ is a consistent estimate of the average treatment effect as long as there is either no effect heterogeneity with respect to $X$ or there is no heterogeneous compliance (i.e. the effect of the instrument on the treatment) with respect to $X$. Otherwise it is inconsistent. The reason is that, if we let $\tilde{T}=T-p_0(X)$ and $\tilde{Z}=Z-r_0(X)$, then the population quantity: $\beta_0(X)=\E[\tilde{T}\tilde{Z}\mid X]$
is a function of $X$. If we also have effect heterogeneity, then we are solving for a constant $\hat{\theta}$ that in the limit satisfies: $\E[(\tilde{Y} - \hat{\theta} \tilde{T}) \tilde{Z}] = 0$,
where $\tilde{Y}=Y - q_0(X)$. On the other hand the true heterogeneous model satisfies the equation: $\E[(\tilde{Y} - \theta_0(X) \tilde{T}) \tilde{Z}] = 0$.
In the limit, the two quantities are related via the equation: $\hat{\theta}\, \E[\tilde{T}\tilde{Z}] = \E[\theta_0(X) \tilde{T} \tilde{Z}]$.
Then the constant effect that we estimate converges to the quantity:
$\hat{\theta} = \frac{\E[\theta_0(X) \beta_0(X)]}{\E[\beta_0(X)]}$.
If $\theta_0(X)$ is not independent with $\beta_0(X)$, then $\hat{\theta}$ is a re-weighted version of the true average treatment effect $\E[\theta(X)]$, re-weighted by the heterogeneous compliance. To account for this heterogeneous compliance we need to change our moment equation so as to re-weight based on $\beta_0(X)$, which is unknown and also needs to be estimated from data.
Given that this function could be arbitrarily complex, we want our final estimate to be robust to estimation errors of $\beta_0(X)$. We can achieve this by considering a doubly robust approach to estimating $\hat{\theta}$. Suppose that we had some other method of computing an estimate of the heterogeneous treatment effect $\theta_0(X)$, then we can combine both estimates to get a more robust method for the ATE, e.g.:
\begin{equation}\label{eqn:dr-ate}
    \textstyle{\hat{\theta}_{DR} = \E\left[\hat{\theta}(X) + \frac{(\tilde{Y} - \hat{\theta}(X)\tilde{T})\tilde{Z}}{\hat{\beta}(X)}\right]}
\end{equation}
This approach has been analyzed in \cite{okui2012doubly} in the case of constant treatment effects and an analogue of this average effect was also used by \cite{athey2017efficient} in a policy learning problem as opposed to an estimation problem. In particular, the quantity $\tilde{Z}/\beta(X)$ is known as the compliance score \cite{ABADIE2003231,aronow2013beyond}. Our methodological contribution in the next two sections is two-fold: i) first we propose a model-based stable approach for estimating a preliminary estimate $\hat{\theta}(X)$, which does not necessarily require that $\beta(X)>0$ everywhere (an assumption that is implicit in the latter method), ii) second we show that this doubly robust quantity can be used as a regression target and minimizing the square loss with respect to this target, corresponds to an orthogonal loss, as defined in \cite{chernozhukov2018plug,foster2019orthogonal}. 
\vsdelete{In fact we show that this loss is orthogonal, even if the hypothesis space that we use in the final model does not contain the true effect function.}

\section{Preliminary Estimate of Conditional Average Treatment Effect (CATE)}

Let $h_0(Z, X) = \E[T\mid Z, X]$ and $p_0$, $q_0$ as in the previous section. Then observe that we can re-write the moment condition as: 
\begin{equation}
    \E[Y - \theta_0(X)\, h_0(Z, X) - f_0(X) \mid Z, X]=0.
\end{equation}
Moreover, observe that the functions $p_0, q_0$ and $f_0$ are related via: $q_0(X) = \theta_0(X)\, p_0(X) + f_0(X)$. Thus we can further re-write the moment condition in terms of $q_0, p_0$ instead of $f_0$:
\begin{equation}
    \E[ Y - q_0(X) - \theta_0(X)\, \left(h_0(Z, X) - p_0(X)\right) \mid Z, X] = 0.
\end{equation}
Moreover, we can identify $\theta(X)$ with the following subset of conditional moments, where the conditioning of $Z$ is removed:
\begin{equation}
\E[ \left(Y - q_0(X) - \theta(X)\, \left(h_0(Z, X) - p_0(X)\right)\right) \, \left(h_0(Z, X) - p_0(X)\right) \mid X] = 0.
\end{equation}
Equivalently, $\theta(X)$ is a minimizer of the square loss:
\begin{equation}
\textstyle{ L^1(\theta; q_0, h_0, p_0) := \E\left[ \left(Y - q_0(X) - \theta(X)\, \left(h_0(Z, X) - p_0(X)\right)\right)^2 \right]}
\end{equation}
since the derivative of this loss with respect to $\theta(X)$ is equal to the moment equation and, thus, the first order condition for the loss minimization problem is satisfied by the true model $\theta_0$. Moreover, if the loss function satisfies a functional analogue of strong convexity, then any minimizer of the loss achieves small mean squared error with respect to $\theta_0$. This leads to the following approach:

\begin{algorithm}[H]
\small
Split the data in half $S_1, S_2$\;
Regress $Y$ on $X$ to learn estimate $\hat{q}$ of function $q_0(X) = \E[Y \mid X]$ on $S_1$\;
Regress $T$ on $Z, X$ to learn estimate $\hat{h}$ of function $h_0(Z, X) = \E[T\mid Z, X]$ on $S_1$\;
Regress $T$ on $X$ to learn estimate $\hat{p}$ of function $p_0(X) = \E[T\mid X]$ on $S_1$\;
Minimize the empirical analogue of the square loss over some hypothesis space $\Theta$ on the other half-sample $S_2$:
\begin{equation}
    \textstyle{\hat{\theta} = \arginf_{\theta\in \Theta} \frac{2}{n}\sum_{i\in S_2} (Y_i - \hat{q}(X_i) - \theta(X_i)\, (\hat{h}(Z_i, X_i) - \hat{p}(X_i))\moedit{)}^2 := L_n^1(\theta; \hat{q}, \hat{h}, \hat{p})}
\end{equation}
or any learning algorithm that achieves small generalization error w.r.t. loss $L^1(\theta; \hat{q}, \hat{h}, \hat{p})$ over $\Theta$.
\caption{\small{{\sc Heterogeneous Effects: DMLIV} Partially orthogonal, convex loss.}}
\label{algo-1:hetero}
\end{algorithm}

This method is an extension of the classical two-stage-least-squares (2SLS) approach \cite{angrist2008mostly} to allow for arbitrary machine learning models; ignoring the residualization part (i.e. if for instance $q(X)=p(X)=0$), then it boils down to: 1) predict the mean treatment from the instrument and $X$ with an arbritrary regression/classification method, 2) predict the outcome from the predicted treatment multiplied by the heterogeneous effect model $\theta(X)$. Residualization helps us remove the dependence of the mean squared error on the complexity of the baseline function $f_0(X)$. We achieve this by showing that this loss is orthogonal with respect to $p, q$ (see \cite{foster2019orthogonal} for the definition of an orthogonal loss). 
However, orthogonality does not hold with respect to $h$. This finding is reasonable since we are using $h(Z, X)$ as our regressor. Hence, any error in the measurement of the regressor can directly propagate to an error in $\theta(X)$. This is the same reason why in classical IV regression one cannot ignore the variance from the first stage of 2SLS when calculating confidence intervals. 

\begin{lemma}\label{thm:orthogonality}
The loss function $L^1(\theta; q, h, p)$ is orthogonal to the nuisance functions $p, q$, but not $h$.
\end{lemma}

\textbf{Strong convexity and overlap.} Note that both the empirical loss $L_n^1$ and the population loss $L^1$ are convex in the prediction, which typically implies computational stability. Moreover, the second order directional derivative of the population loss in any functional direction $\theta(\cdot) - \theta_0(\cdot)$ is:
\begin{equation}
    \E\left[(\hat{h}(Z, X) - \hat{p}(X))^2\, \left(\theta(X)-\theta_0(X)\right)^2\right]
\end{equation}
and let: 
\begin{equation}
    V(X):=\E\left[(\hat{h}(Z, X) - \hat{p}(X))^2\mid X\right]
\end{equation}
To be able to achieve mean-squared-error rates based on our loss minimization, we need the population version $L^1$ of the loss function to satisfy a functional analogue of $\lambda$-strong convexity:
\begin{equation}
    \forall \theta \in \Theta: \E[V(X) \cdot (\theta(X) - \theta_0(X))^2] \geq \lambda\, \E[(\theta(X)-\theta_0(X))^2]
\end{equation}
This setting falls under the ``single-index'' setup of \cite{foster2019orthogonal}. Using arguments from Lemma~1 of \cite{foster2019orthogonal}, if:
\begin{equation}
    \forall \theta \in \Theta: \E[V_0(X) \cdot (\theta(X) - \theta_0(X))^2] \geq \lambda_0\, \E[(\theta(X)-\theta_0(X))^2]
\end{equation}
where
\begin{equation}
    V_0(X):=\E\left[(h_0(Z, X) - p_0(X))^2\mid X\right] = \Var(\E[T\mid Z, X] \mid X),
\end{equation}
then $\lambda\geq \lambda_0 - O(\|h-h_0\|_4^2, \|p-p_0\|_4^2)=\lambda_0-o(1)$.
A sufficient condition is that $V_0(X)\geq \lambda_0$ for all $X$. This is a standard "overlap" condition that the instrument is exogenously varying at any $X$ and has a direct effect on the treatment at any $X$. 
DMLIV only requires an "average" overlap condition, tailored particularly to the hypothesis space $\Theta$, hence it could handle settings where the instrument is weak for some subset of the population.
For instance, if $\Theta$ is a linear function class: $\Theta=\{\ldot{\theta}{\phi(X)}: \theta \in S \subseteq \R^d\}$, then for the oracle strong convexity to hold it suffices that:
$\E[V_0(X) \phi(X)\phi(X)^T]\succeq \lambda I$.
Lemma~\ref{thm:orthogonality}, combined with the above discussion and the results of \cite{foster2019orthogonal} yields:\footnote{This corollary follows by small modifications of the proofs of Theorem~1 and  Theorem~3 of \cite{foster2019orthogonal} that accounts for the non-orthogonality w.r.t. $h$, so we omit its proof.}
\begin{corollary}\label{cor:dmliv}
Assume all random variables are bounded. Suppose that in the final stage of DMLIV we use any algorithm that achieves expected generalization error $R_n^2$ w.r.t. loss $L^1(\theta; \hat{q}, \hat{h}, \hat{p})$, i.e.:
\begin{equation}
    \E\left[L^1(\hat{\theta}_{DR}; \hat{q}, \hat{h}, \hat{p}) - \inf_{\theta\in \Theta} L^1(\theta; \hat{q}, \hat{h}, \hat{p})\right] \leq R_n^2
\end{equation}
Moreover, suppose that the nuisance estimates satisfy $\E[\|\hat{q}-q_0\|_4^4],\E[\|\hat{p}-p_0\|_4^4]=o(g_n^4)$ and $\E[\|\hat{h}-h_0\|_4^4]=o(h_n^4)$ and for all $\theta\in \Theta$: $\E[V_0(X) \cdot (\theta(X) - \theta_0(X))^2] \geq \lambda_0\, \E[(\theta(X)-\theta_0(X))^2]$. Then $\hat{\theta}$ returned by DMLIV satisfies:
\begin{equation}
    \E\left[\left(\hat{\theta}(X)-\theta_0(X)\right)^2\right] \leq O\left(\frac{R_n^2 + h_n^2 + g_n^4}{\lambda_0}\right)
\end{equation}
If empirical risk minimization is used in the final stage, i.e.:
\begin{equation}
\hat{\theta} = \arg\min_{\theta \in \Theta} L_n^1(\theta; \hat{q}, \hat{h}, \hat{p})
\end{equation} then $R_n^2=\delta_n^2 + h_n^2 + g_n^4$, where $\delta_n$ is the critical radius of the hypothesis space $\Theta$ as defined via the localized Rademacher complexity \cite{koltchinskii2011introduction}.
\end{corollary}
\vsdelete{We note that due to the non-orthogonality of the loss w.r.t. to $h$, the error in $h$ appears in the same order in the final MSE rate. However, since we are only using DMLIV as a preliminary estimator, this result is good enough for our purposes of achieving a preliminary MSE rate.}
\textbf{Computational considerations.} The empirical loss $L_n^1$ is not a standard square loss.
However, we can re-write it as $\sum_i \gamma(X_i)^2 (\tilde{Y}_i/\gamma(X_i) - \theta(X_i))^2$. Thus the problem is equivalent to a standard square loss minimization with label $\tilde{Y}_i/\gamma(X_i)$ and sample weights $\gamma(X_i)^2$. Thus we can use any out-of-the-box machine learning method that accepts sample weights, such as stochastic gradient based regression methods and gradient boosted or random forests. Alternatively, if we assume a linear representation of the effect function $\theta(X)=\ldot{\theta}{\phi(X)}$, then the problem is equivalent to regressing $\tilde{Y}$ on the scaled features $\phi(X)\, \gamma(X)$, and again any method for fitting linear models can be invoked.

\section{DRIV: Orthogonal Loss for IV Estimation of CATE and Projections}

We now present the main estimation algorithm that combines the doubly robust approach presented for ATE estimation with the preliminary estimator of the CATE to obtain a fully orthogonal and strongly convex loss. \moedit{This method achieves a} second order effect from all nuisance estimation errors and \moedit{enables} oracle rates for the target effect class $\Theta$ and asymptotically valid inference for low dimensional target effect classes. 
In particular, given access to a first stage model of heterogeneous effects $\theta_{pre}$ (such as the one produced by DMLIV), we can estimate a more robust model of heterogeneous effects via minimizing a square loss that treats the doubly robust quantity used in Equation~\eqref{eqn:dr-ate} as the label:
\begin{equation}
    \textstyle{\min_{\theta\in \Theta_{\pi}} L^2(\theta; \theta_{pre}, \beta, p, q, r) :=  \E\left[ \left(\theta_{pre}(X) + \frac{(\tilde{Y} - \theta_{pre}(X)\tilde{T})\tilde{Z}}{\beta(X)} - \theta(X)\right)^2\right]}
\end{equation}
We 
allow for a model space $\Theta_{\pi}$ that is not necessarily equal to $\Theta$. The solution in Equation~\eqref{eqn:dr-ate} is a special case of this minimization problem where the space $\Theta_{\pi}$ contains only constant functions.
Our main result \moedit{shows} that this loss is orthogonal to all nuisance functions $\theta_{pre}$, $\hat{\beta}$, $\hat{q}$, $\hat{p}$, $\hat{r}$. Moreover, it is strongly convex in the prediction $\theta(X)$, since conditional on all the nuisance estimates it is a standard square loss. Moreover, we show that the loss is orthogonal irrespective of what the model space $\Theta_{\pi}$, even if $\Theta_{\pi}\neq \Theta$, as long as the preliminary estimate $\theta_{pre}$ is consistent with respect to the true CATE $\theta_0$ (i.e. fit a flexible preliminary CATE and use it to project to a simpler hypothesis space). 

\begin{lemma}\label{thm:dr-orthogonality}
    The loss $L^2$ is orthogonal with respect to the nuisance functions $\theta_{pre}$, $\beta$, $p$, $q$ and $r$.
\end{lemma}

\vsdelete{This leads to the following algorithm for heterogeneous treatment effects:}

\begin{algorithm}[H]
\small
Estimate a preliminary estimate $\theta_{pre}$ of the CATE $\theta_0(X)$ using DMLIV on half-sample $S_1$\;
Using half-sample $S_1$, regress i) $Y$ on $X$, ii) $T$ on $X$, iii) $Z$ on $X$ to learn estimates $\hat{q}, \hat{p}, \hat{r}$ correspondingly\;
Regress $T\cdot Z$ on $X$ using $S_1$ to learn estimate $\hat{f}$ of function $f_0(X)=\E[T\cdot Z\mid X]$\;
$\forall i\in S_2$, let $\tilde{Y}_i = Y_i - \hat{q}(X_i)$, $\tilde{T}_i = T_i - \hat{p}(X_i)$, $\tilde{Z}_i = Z_i - \hat{r}(X_i)$, $\hat{\beta}(X_i) = \hat{f}(X_i) - \hat{p}(X_i)\, \hat{r}(X_i)$\;
Minimize empirical analogue of square loss $L^2$ over hypothesis space $\Theta_{\pi}$ on the other half-sample $S_2$,
i.e.:
\begin{equation*}
    \textstyle{\hat{\theta}_{DR} = \arginf_{\theta\in \Theta_{\pi}} \frac{2}{n}\sum_{i\in S_2} \left(\theta_{pre}(X_i) + \frac{(\tilde{Y}_i - \theta_{pre}(X_i)\, \tilde{T}_i)\tilde{Z}_i}{\hat{\beta}(X_i)} - \theta(X_i)\right)^2:=L_n^2(\theta; \theta_{pre}, \hat{\beta}, \hat{p}, \hat{q}, \hat{r})}
\end{equation*}
or any learning algorithm that has small generalization error w.r.t. loss $L^2(\theta; \theta_{pre}, \hat{\beta}, \hat{p}, \hat{q}, \hat{r})$ on $\Theta_{\pi}$.
\caption{\small{{\sc DRIV} Orthogonal convex loss for CATE and projections of CATE}}
\label{algo-1:hetero}
\end{algorithm}

If we use DMLIV for $\theta_{prel}$, even though DMLIV has a first order impact from the error of $h$, the second stage estimate has a second order impact, since it has a second order impact from the first stage CATE error.
Lemma~\ref{thm:dr-orthogonality} together with the results of \cite{foster2019orthogonal} implies the following corollary:
\begin{corollary}\label{cor:driv}
Assume all random variables are bounded and $\beta(X)\geq \beta_{\min} >0$ for all $X$. Suppose that in the final stage of DRIV we use any algorithm that achieves expected generalization error $R_n^2$ with respect to loss $L^2(\theta; \theta_{pre}, \hat{\beta}, \hat{p}, \hat{q}, \hat{r})$ over hypothesis space $\Theta_{\pi}$, i.e.:
\begin{equation}
    \E\left[L^2(\hat{\theta}_{DR}; \theta_{pre}, \hat{\beta}, \hat{p}, \hat{q}, \hat{r}) - \inf_{\theta\in \Theta} L^2(\theta; \theta_{pre}, \hat{\beta}, \hat{p}, \hat{q}, \hat{r})\right] \leq R_n^2
\end{equation}
Moreover, suppose that each nuisance estimate $\hat{g}\in \{\theta_{pre}, \hat{\beta}, \hat{p}, \hat{q}, \hat{r}\}$, $\E[\|\hat{g}-g_0\|_4^4]\leq g_n^4$ and the hypothesis space $\Theta_{\pi}$ is convex. Then $\hat{\theta}_{DR}$ returned by DRIV satisfies: 
\begin{equation}
\E[\|\hat{\theta}_{DR}-\theta_*\|_2^2] \leq O\left(R_n^2 + g_n^4\right),
\end{equation}
where 
\begin{equation}
    \theta_*=\arg\min_{\theta \in \Theta_{\pi}} L^2(\theta; \theta_0, \beta_0, p_0, q_0, r_0) \equiv \arg\min_{\theta \in \Theta_{\pi}}  \E[\left(\theta_0(X) - \theta(X)\right)^2]
\end{equation}
If empirical risk minimization is used in the final stage, i.e.:
\begin{equation}
\hat{\theta}_{DR} = \arg\min_{\theta \in \Theta_{\pi}} L_n^2(\theta; \theta_{pre}, \hat{\beta}, \hat{p}, \hat{q}, \hat{r})
\end{equation}
then $R_n^2=\delta_n^2 + g_n^4$, where $\delta_n$ is the critical radius of the hypothesis space $\Theta$ as defined via the localized Rademacher complexity \cite{koltchinskii2011introduction}. 
\end{corollary}
For the special case where $\Theta_{\pi}$ is high-dimensional sparse linear and contains the true CATE, then it suffices to require only an $\|\cdot\|_2$ rate for the nuisance functions, as opposed to an $\|\cdot\|_4$. This corollary stems from verifying that all the assumptions of \cite{chernozhukov2018plug} hold for our loss function:
\begin{corollary}
If $\Theta_{\pi}$ is high-dimensional sparse linear, i.e. $\theta(X)=\ldot{\xi}{\phi(X)}$ with $\|\xi\|_0\leq s$, $\phi(X)\in \R^p$ and $\E[\phi(X)\phi(X)^T]\geq \gamma_0 I$, and the model is well-specified, i.e. $\theta_0\in \Theta_{\pi}$, with $\theta_0=\ldot{\xi_0}{\phi(X)}$, then if $\ell_1$-penalized square loss minimization is used in the final step of DRIV, i.e.:
\begin{equation}
\hat{\theta}_{DR} = \arg\min_{\xi \in \R^p} L_n^2(\xi; \theta_{pre}, \hat{\beta}, \hat{p}, \hat{q}, \hat{r}) + \lambda \|\xi\|_1
\end{equation}
it suffices that $\E[\|\hat{g}-g_0\|_2^2]\leq g_n^2$ to get: $\E[\|\hat{\xi}-\xi_*\|_2^2] \leq O\left(s^2\frac{\log(p)/n+g_n^4}{\gamma_0}\right)$ for $\lambda = \Theta(1/\sqrt{n})$.
\end{corollary}


\textbf{Interpretability through projections.}
The fact that our loss function can be used with any target $\Theta_{\pi}$ allows us to perform inference on the projection of $\theta_0$ on a simple space $\Theta_{\pi}$ (e.g. decision trees, linear functions) for interpretability purposes. 
If we let $Y_i^{DR}$ the label in the final regression of DRIV, then observe that when the nuisance estimates take their true values then $\E[Y_i^{DR}\mid X]=\theta_0(X)$, since the second part of $Y_i^{DR}$ has mean zero. Hence:
\begin{equation}
    \textstyle{L^2(\theta; \theta_0, \beta_0, p_0, q_0, r_0) = \Var(\theta_{DR}(X)) + \E[\left(\theta_0(X) - \theta(X)\right)^2]}.
\end{equation}
The first part is independent of $\theta$ and hence minimizing the oracle $L^2$ is equivalent to minimizing $\E[\left(\theta_0(X) - \theta(X)\right)^2]$ over $\theta\in \Theta_{\pi}$, which is exactly the projection of $\theta_0$ on $\Theta_{\pi}$. 
One version of an interpretable model is estimating the CATE with respect to a subset $T$ of the variables, i.e.: $\theta(X_T) = \E[\theta_0(X)\mid X_T]$ (e.g. how treatment effect varies with a single feature). This boils down to setting $\Theta_{\pi}$ some space of functions of $X_T$.

If $T$ is a low dimensional set of features and $\Theta_{\pi}$ is a 
the space of linear functions of $X_T$,
i.e. $\Theta_{\pi}=\{X\rightarrow \ldot{\theta_T}{X_T}: \theta_T\in \R^{|T|}\}$, then the first order condition of our loss 
is equal to the moment condition $\E[(Y^{DR}-\ldot{\theta_T}{X_T})X_T]=0$. Then orthogonality of our loss implies that DRIV is equivalent to an orthogonal moment estimation method \cite{Chernozhukov2018}. Thus using the results of \cite{Chernozhukov2018} we get the corollary:
\begin{corollary}[Confidence Intervals]
The estimate $\hat{\theta}_T$ returned by DRIV with $\Theta_{\pi}=\{X\rightarrow \ldot{\theta_T}{X_T}: \theta_T\in \R^{|T|}\}$ for $|T|$ a constant independent of $n$, is asymptotically normal with asymptotic variance equal to the hypothetical variance of $\theta_T$ as if the nuisance estimates had their true values.
\end{corollary}
Hence, we can use out-of-the-box packages for calculating CIs of an OLS regression to get $p$-values on the coefficients.


\subsection{Variance Reduction with Re-Weighting under Well-Specification}

DRIV is orthogonal irrespective of whether $\Theta_{\pi}$ contains the true CATE model and provides estimation rates for the projection of CATE on the space $\Theta_{\pi}$. However, it does suffer from a high variance, since we are dividing by the conditional co-variance $\beta(X)$. Hence, if the instrument is weak in some region of $X$'s then the method can suffer from instability. Moreover, the variance of the approach does not adapt to the model space $\Theta_{\pi}$, i.e. it could be that some $X$'s have a zero co-variance but the model $\theta_0(X)$ is identified by the remainder of the $X$'s. 

Unlike DRIV, DMLIV does not suffer from these drawbacks. However, as pointed out earlier, DMLIV is not orthogonal with respect to the nuisance function $h(x,z) = E[T|X=x, Z=z]$. We now show how we can combine the best of both worlds by simply re-weighting the DRIV loss, so as to put less weight on high variance regions. However, unlike DRIV, the loss that we construct is simply orthogonal and not universally orthogonal. In particular, since $1/\beta(X)^2$ is a reasonable proxy on the magnitude of the variance of the regression target $Y_{DR}$, we will re-weight the loss by $\beta(X)^2$:
\begin{equation}
    L_{rw}^2(\theta; \theta_{pre}, \beta, p, q, r)=\E\left[\left(\tilde{Y} \tilde{Z} - \theta_{pre}(X)\,(\tilde{T}\, \tilde{Z} - \beta(X)) + \theta(X)\, \beta(X)\right)^2\right]
\end{equation}
In other words we are re-weighting the DRIV loss to give less weight on samples that have high variance and solely using the samples with low variance to identify our true model. We can then show the following theorem:
\begin{lemma}\label{thm:dr-orthogonality-well-spec}
    Assuming $\theta_0\in \Theta_{\pi}$, then the loss $L_{rw}^2(\theta; \theta_{pre}, \beta, p, q, r)$ is orthogonal with respect to the nuisance functions $\theta_{pre}$, $\beta$, $p$, $q$ and $r$. 
\end{lemma}
We will refer to latter algorithm as DRIV-RW (for re-weighted DRIV). Similar to DRIV we can then get the following corollary:
\begin{corollary}\label{cor:driv-rw}
Assume all random variables are bounded, $\theta_0\in \Theta_{\pi}$ and suppose that in the final stage of DRIV-RW we use any algorithm that achieves expected generalization error $R_n^2$ with respect to loss $L_{rw}^2(\theta; \theta_{pre}, \hat{\beta}, \hat{p}, \hat{q}, \hat{r})$ over hypothesis space $\Theta_{\pi}$, i.e.:
\begin{equation}
    \E\left[L_{rw}^2(\hat{\theta}_{DR}; \theta_{pre}, \hat{\beta}, \hat{p}, \hat{q}, \hat{r}) - \inf_{\theta\in \Theta} L_{rw}^2(\theta; \theta_{pre}, \hat{\beta}, \hat{p}, \hat{q}, \hat{r})\right] \leq R_n^2
\end{equation}
Moreover, suppose that each nuisance estimate $\hat{g}\in \{\theta_{pre}, \hat{\beta}, \hat{p}, \hat{q}, \hat{r}\}$, $\E[\|\hat{g}-g_0\|_4^4]\leq g_n^4$ and for all $\theta\in \Theta$: $\E[\beta_0(X)^2 \cdot (\theta(X) - \theta_0(X))^2] \geq \lambda_0\, \E[(\theta(X)-\theta_0(X))^2]$. Then $\hat{\theta}_{DR}$ returned by DRIV-RW satisfies: 
\begin{equation}
    \E[(\hat{\theta}_{DR}(X) - \theta_0(X))^2]  \leq O\left(\frac{R_n^2 + g_n^4}{\lambda_0}\right),
\end{equation}
\end{corollary}
We can also make statements analogous to Corollary~\ref{cor:driv} for the case where empirical risk minimization is used in the final stage or $\ell_1$-penalized empirical risk minimization over linear functions. Moreover, observe that even if the on-average overlap condition $\E[\beta_0(X)^2 \cdot (\theta(X) - \theta_0(X))^2] \geq \lambda_0\, \E[(\theta(X)-\theta_0(X))^2]$ does not hold, we are still recovering a model that converges to the true one in terms of the re-weighted MSE, where we re-weight based on a measure of strength of the instrument (i.e. we will be predicting better in regions where the instrument is strong). This is a good fall-back guarantee to have in cases where the instrument happens to be weak.

Finally, in the case where the instrument $Z$ is multi-dimensional, then we can follow an approach similar to DMLIV, where we construct an ``optimal'' instrument of the form $Z_{\pi}=\E[T\mid Z, X]$. Then we can consider the analogue of DRIV-RW, but where $Z_{\pi}$ takes the place of $Z$, i.e. let:
\begin{equation}
    \tilde{Z}_{\pi} = h(X, Z)-p(X)
\end{equation}
Then observe that the analogue of $\beta$ for the instrument $Z_{\pi}$ is equal to $V(X)$ as defined in the DMLIV section, i.e.
\begin{equation*}
    \beta_{\pi}(X) = \E[\tilde{T}\,\tilde{Z}_{\pi}\mid X] = \E[\tilde{Z}_{\pi}^2\mid X] = \Var(\E[T\mid Z, X]\mid X) = V(X)
\end{equation*}
then the loss function takes the form:
\begin{equation}
    L_{\pi, rw}^2(\theta; \theta_{pre}, V, p, q, h)=\E\left[\left(\tilde{Y} \tilde{Z}_{\pi} - \theta_{pre}(X)\,(\tilde{T}\, \tilde{Z}_{\pi} - V(X)) + \theta(X)\, V(X)\right)^2\right]
\end{equation}

\begin{lemma}\label{thm:projected-dr-orthogonality}
 Assuming that there exists $\theta_0\in \Theta_{\pi}$, such that:
\begin{equation}
    \E\left[(Y - q_0(X) - \theta_0(X)\, (T - p_0(X)))\, (h_0(Z, X)-r_0(X))\mid X\right]=0
\end{equation}
and that the preliminary estimate $\theta_{pre}$ converges in mean-squared-error to $\theta_0$ then the loss $L_{\pi, rw}^2(\theta; \theta_{pre}, V, p, q, h)$ is orthogonal with respect to the nuisance functions $\theta_{pre}$, $V$, $p$, $q$, $h$. 
\end{lemma}
Observe that if we use DMLIV as the preliminary estimator, then because DMLIV solely uses the latter set of moment restrictions, it will converge in MSE to a function that satisfies the moment condition. Thus it will satisfy the requirement. We will refer to this algorithm as Projected-DRIV-RW.
\begin{corollary}\label{cor:projected-driv-rw}
Assume all random variables are bounded, $\theta_0\in \Theta_{\pi}$ and suppose that in the final stage of Projected-DRIV-RW we use any algorithm that achieves expected generalization error $R_n^2$ with respect to loss $L_{\pi, rw}^2(\theta; \theta_{pre}, \hat{V}, \hat{p}, \hat{q}, \hat{h})$ over hypothesis space $\Theta_{\pi}$, i.e.:
\begin{equation}
    \E\left[L_{\pi, rw}^2(\hat{\theta}_{DR}; \theta_{pre}, \hat{V}, \hat{p}, \hat{q}, \hat{h}) - \inf_{\theta\in \Theta} L_{\pi, rw}^2(\theta; \theta_{pre}, \hat{V}, \hat{p}, \hat{q}, \hat{h})\right] \leq R_n^2
\end{equation}
Moreover, suppose that each nuisance estimate $\hat{g}\in \{\theta_{pre}, \hat{V}, \hat{p}, \hat{q}, \hat{h}\}$, $\E[\|\hat{g}-g_0\|_4^4]\leq g_n^4$ and for all $\theta\in \Theta$: $\E[V_0(X)^2 \cdot (\theta(X) - \theta_0(X))^2] \geq \lambda_0\, \E[(\theta(X)-\theta_0(X))^2]$. Then $\hat{\theta}_{DR}$ returned by Projected-DRIV-RW satisfies: 
\begin{equation}
    \E[(\hat{\theta}_{DR}(X) - \theta_0(X))^2]  \leq O\left(\frac{R_n^2 + g_n^4}{\lambda_0}\right),
\end{equation}
\end{corollary}

\section{Example: Randomized Trials with Non-Compliance}

Given that our main application is a special case of our framework where the instrument is the assignment in a randomized control trial with non-compliance, in this section we show how our algorithms simplify for this setting. 

Randomized control trials with non-compliance, or equivalently intent-to-treat A/B tests, are the special case where the instrument and the treatment are binary, i.e. $Z, T\in \{0,1\}$, and the instrument $Z$ is fully exogenous. For simplicity, we assume that $\Pr[Z=1\mid X]=1/2$ for all $X$ (i.e. a balanced $A/B$ test). In this case, the nuisance components can all be expressed as a function of the following quantity (which is typically referred to as the compliance score):
\begin{align}
    \Delta(X) =~& (2\, Z - 1)\,\frac{\Pr[T=1\mid Z=1, X] - \Pr[T=1\mid Z=0, X]}{2}
\end{align}
Then, after straightforward algebraic manipulations, we have:
\begin{align*}
    h(Z, X) - p(X) =~& \Delta(X)\\
    \beta(X) =~& (2\, Z-1) \frac{\Delta(X)}{2}
\end{align*}
and the loss functions simplify to:
\begin{align*}
    L^1(\theta; q, \Delta) =~& \E\left[\left( Y - q(X) - \theta(X)\, \Delta(X)\right)^2\right]\\
    L^2(\theta; \theta_{pre}, q, p, \Delta) =~& \E\left[\left(\theta_{pre}(X) + \frac{Y-q(X) - \theta_{pre}(X)\, (T-p(X))}{\Delta(X)} - \theta(X)\right)^2\right]\\
    L_{rw}^2(\theta; \theta_{pre}, q, p, \Delta) =~& \frac{1}{4}\E\left[\left(Y-q(X) - \theta_{pre}(X)\, (T-p(X)) + \Delta(X)\, (\theta_{pre}(X)-\theta(X))\right)^2\right]
\end{align*}
Moreover, observe that the loss $L^1$ is equivalent to the loss $L_{rw}^2$ with $\theta_{pre}(X)=0$, i.e. a zero preliminary estimator of the treatment effect. Thus in this case, we essentially need to estimate three nuisance components, i.e. $q, p, \Delta$. We can estimate $\Delta$ by simply estimating $h(Z, X)=\Pr[T=1\mid Z, X]$ and setting $\Delta(X)=(2\,Z-1) (h(1, X)-h(0, X))/2$, which boils down to a classification problem among the treated and control populations for each value of $Z$. In fact, once we have estimated these two nuisance functions, we also know that: $p(X) = \frac{h(1, X) + h(0, X)}{2}$. Thus it all boils down to estimating $q$ and $h$.

Finally, the re-weighted loss, can be further simplified to:
\begin{equation}
    L_{rw}^2(\theta; \theta_{pre}, q, p, \Delta) = \frac{1}{4}\E\left[\left(Y-q(X) - \theta_{pre}(X)\, (T-h(Z, X)) - \Delta(X)\, \theta(X)\right)^2\right]
\end{equation}
This functional form is reminiscent of the R-learner loss for the case when the treatment is exogenous, which corresponds to: $E[(Y-q(X) - \theta(X)\, (T-p(X))^2]$. Here, we see that to handle the endogeneity of the treatment, we need to fit a preliminary estimator.

\section{Estimating Effects of Membership on TripAdvisor}

We apply our methods to estimate the treatment effect of membership on the number of days a user visits a leading travel assistance website, W. The instrument used was a 14-day intent-to-treat A/B test run during 2018, where users in group A received a new, easier membership sign-up process, while the users in group B did not. The treatment is whether a user became a member or not. Becoming a member and logging into TripAdvisor gives users exclusive access to trip planning tools, special deals and price alerts, and personalized ideas and travel advice. 
Our data consists of 4,606,041 total users in a 50:50 A/B test. For each user, we have a 28-day pre-experiment summary about their browsing and purchasing activity on TripAdvisor (see Sec. \ref{sec:Wapp-additional-data}). The instrument significantly increased the rate of treatment, and is assumed to satisfy the exclusion restriction.

\begin{figure}[htpb]
\begin{subfigure}{.26\textwidth}
    \centering
    \includegraphics[width=.95\linewidth]{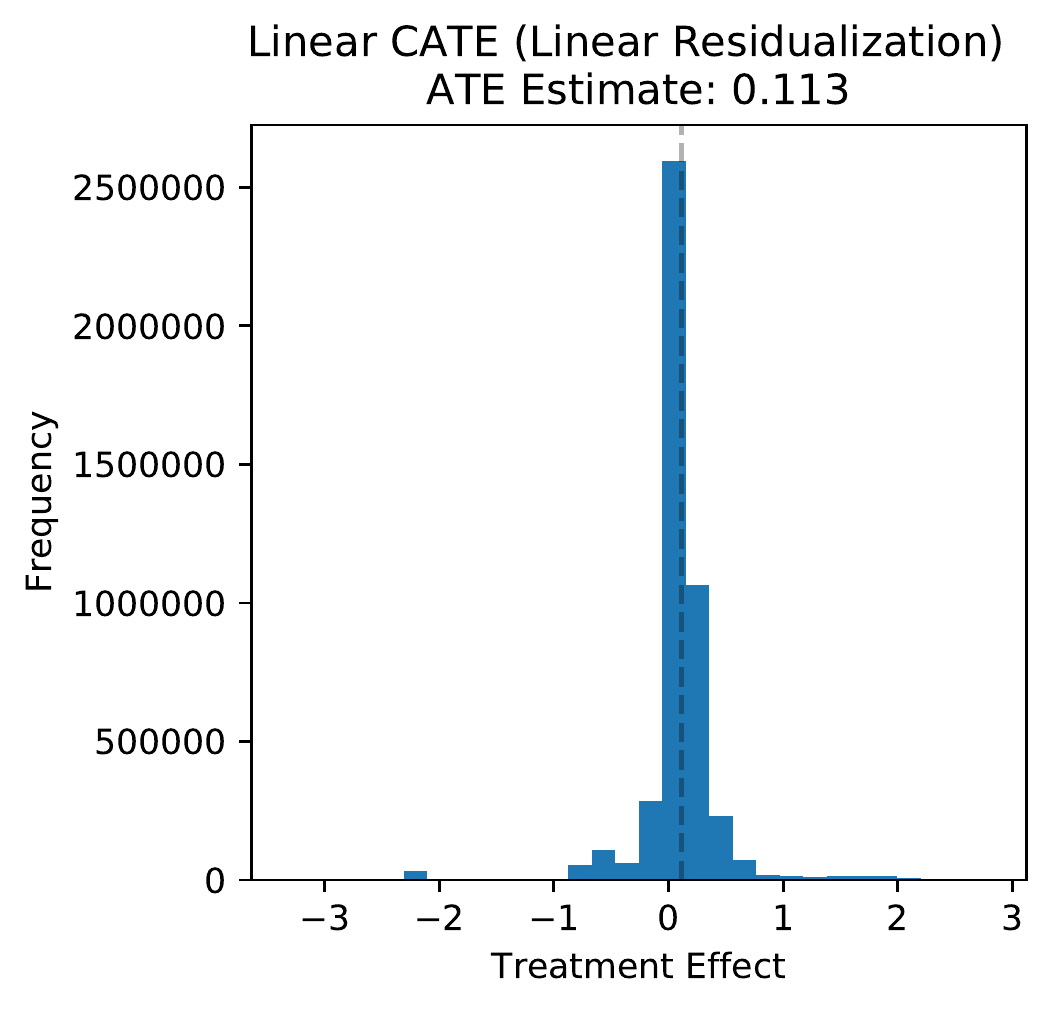}
    \label{fig:linear-cate-w-linear-resid_norm}
\end{subfigure}%
\begin{subfigure}{.37\textwidth}
    \centering
    \includegraphics[width=\linewidth]{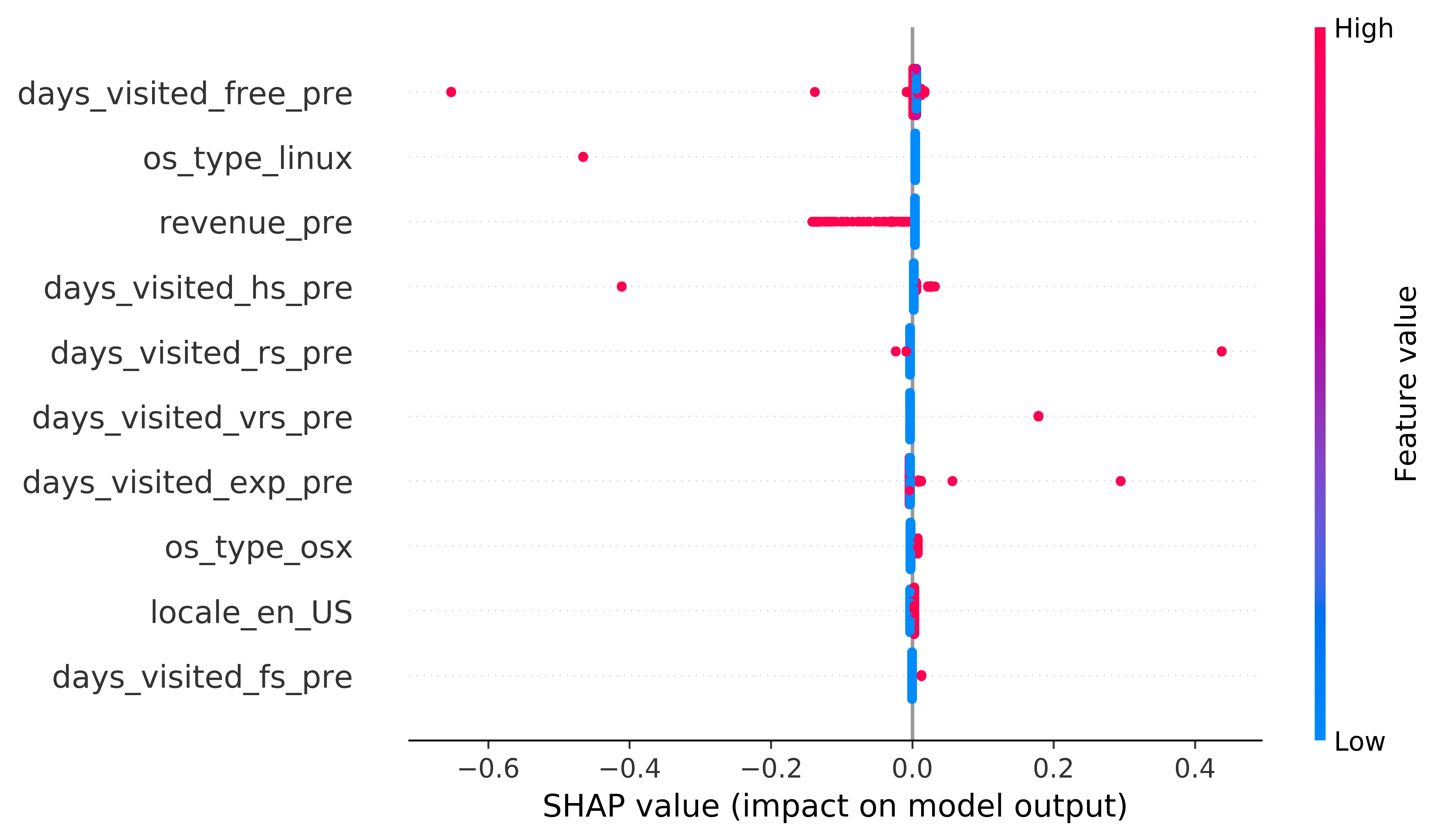}
    \label{fig:linear-cate-w-linear-resid_norm}
\end{subfigure}%
\begin{subfigure}{.37\textwidth}
    \centering
    \includegraphics[width=.95\linewidth]{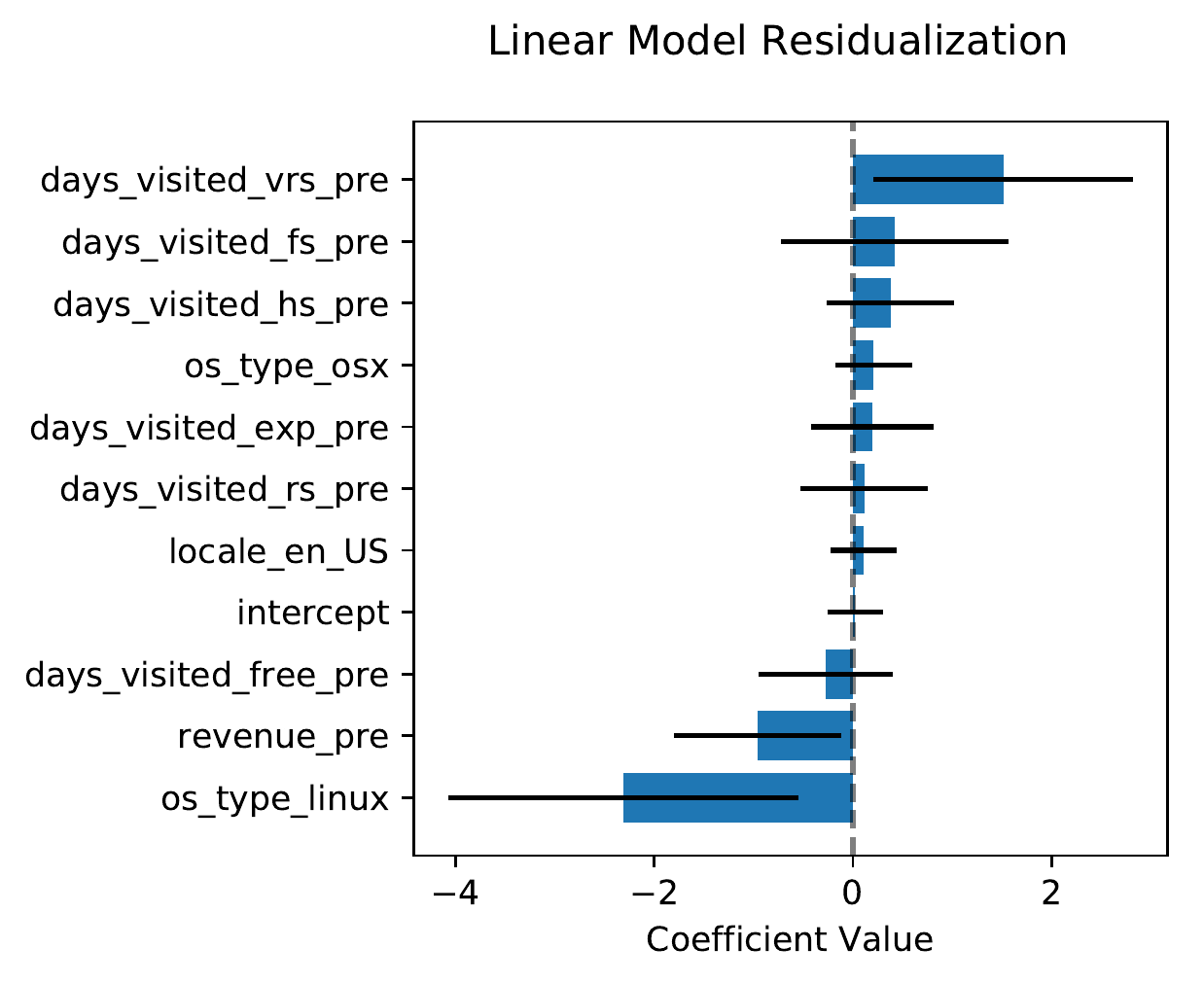}
    \label{fig:linear-cate-w-linear-resid_norm}
\end{subfigure}
\caption{\small{(From left to right) Linear CATE projection, SHAP summary of random forest CATE projection, Linear CATE projection coefficients. Using linear nuisance models.}}
\label{fig:3sub-linear-TA}
\end{figure}

We applied two sets of nuisance estimation models with different complexity characteristics: LASSO regression and logistic regression with an L2 penalty (LM); and gradient boosting regression and classification (GB). The only exception was $E[Z|X]$, where we used a fixed estimate of $0.5$ since the instrument was a large randomized experiment. See Sec. \ref{sec:Wapp-model-details} for details.\footnote{We attempted to use the R implementation of Generalized Random Forests (GRF)\cite{athey2019generalized} to compare with our results. However, we could not fit due to the size of the data and insufficient memory errors (with 64GB RAM). 
}
\begin{table}[H]
\footnotesize
\centering
\small
\begin{minipage}{.5\linewidth}
\begin{tabular}{c c c c} 
 \hline
 Nuisance & Method & ATE Est [95\% CI] \\
 \hline
 LM & DMLATEIV & 0.117 [-0.051, 0.285] \\
 LM &  DRIV & 0.113 [-0.052, 0.279] \\
\end{tabular}
\end{minipage}%
\begin{minipage}{.5\linewidth}
\begin{tabular}{c c c c} 
 \hline
 Nuisance & Method & ATE Est [95\% CI] \\
 \hline
 GB & DMLATEIV & 0.127 [-0.031, 0.285] \\
 GB & DRIV & 0.125 [-0.061, 0.311] \\
\end{tabular}
\end{minipage}%
\caption{ATE Estimates for 2018 Experiment at W}
\label{table:W_ate_compare}
\end{table}
We estimate the ATE using DRIV projected onto a constant (Table \ref{table:W_ate_compare}). Using linear nuisance models results in very similar ATE estimates between DMLATEIV and DRIV. We compare the $X$ co-variate associations for both heterogeneity and compliance under DRIV to understand why. If there are co-variates with significant non-zero associations in both heterogeneity and compliance, this could lead to different estimates between DRIV and DMLATEIV (and vice versa).
Replacing the CATE projection model with a linear regression, we obtain valid inferences for the co-variates associated with treatment effect heterogeneity (Figure \ref{fig:3sub-linear-TA}). For compliance, we run a linear regression of the estimated quantity $\beta(X)$ on $X$, to assess its association with each of the features (see Sec. \ref{sec:Wapp-model-details} for details). Comparing treatment and compliance coefficients, os\_type\_linux and revenue\_pre are the only coefficients substantially different from 0 in both. However, only a very small proportion of users in the experiment are Linux users, and the distribution of revenue is very positively skewed. This justifies the minor difference between the DMLATEIV and DRIV estimates.
Moreover, we fit a shallow, heavily regularized random forest and interpret it using Shapley Additive Explanations (SHAP) \cite{NIPS2017_7062}. \vsedit{SHAP gave directionally similar impact of each feature on the effect} 
(Figure \ref{fig:3sub-linear-TA}). However, since we constrained the model to have depth at most one, it essentially gives the features in order of importance if we were to split the population based on a single feature. This justifies why the order of importance of features in the forest is not in the same order as the magnitude of the rank in the linear model, since they have different interpretations. The features picked up by the forest intuitively make sense since an already highly engaged member of W, or a user who has recently made a booking, is less likely to further increase their visits to W. 

\begin{figure}[htpb]
\begin{subfigure}{.26\textwidth}
    \centering
    \includegraphics[width=.95\linewidth]{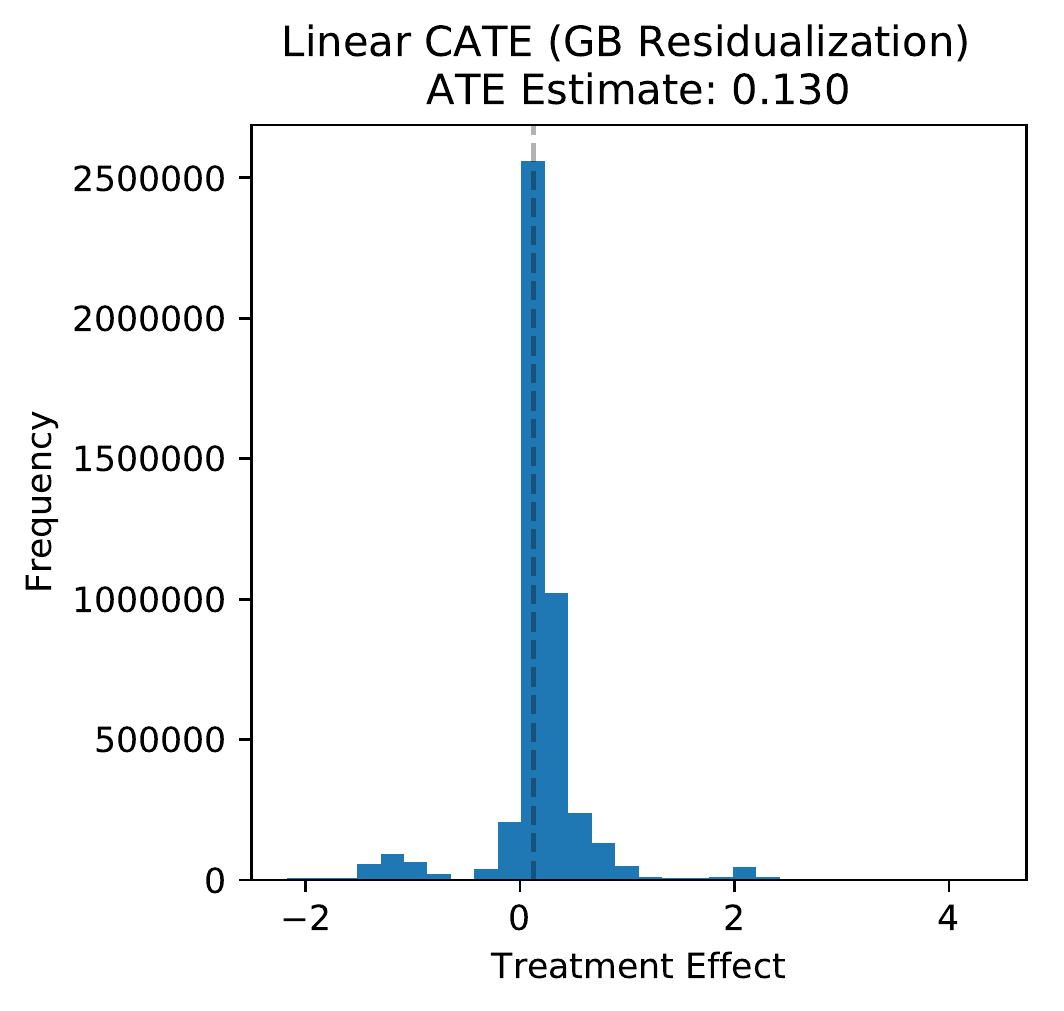}
\end{subfigure}%
\begin{subfigure}{.37\textwidth}
    \centering
    \includegraphics[width=\linewidth]{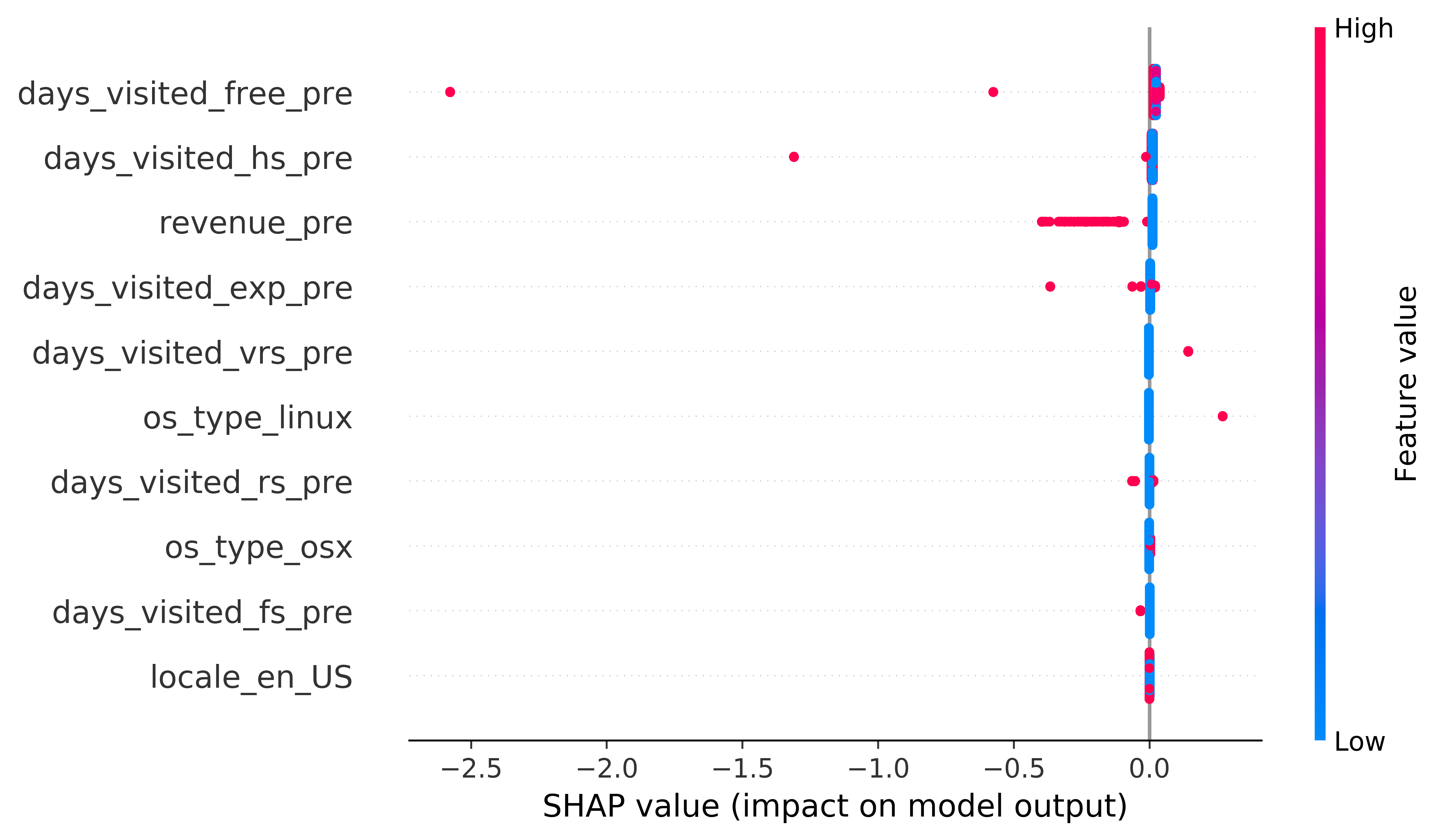}
\end{subfigure}%
\begin{subfigure}{.37\textwidth}
    \centering
    \includegraphics[width=.95\linewidth]{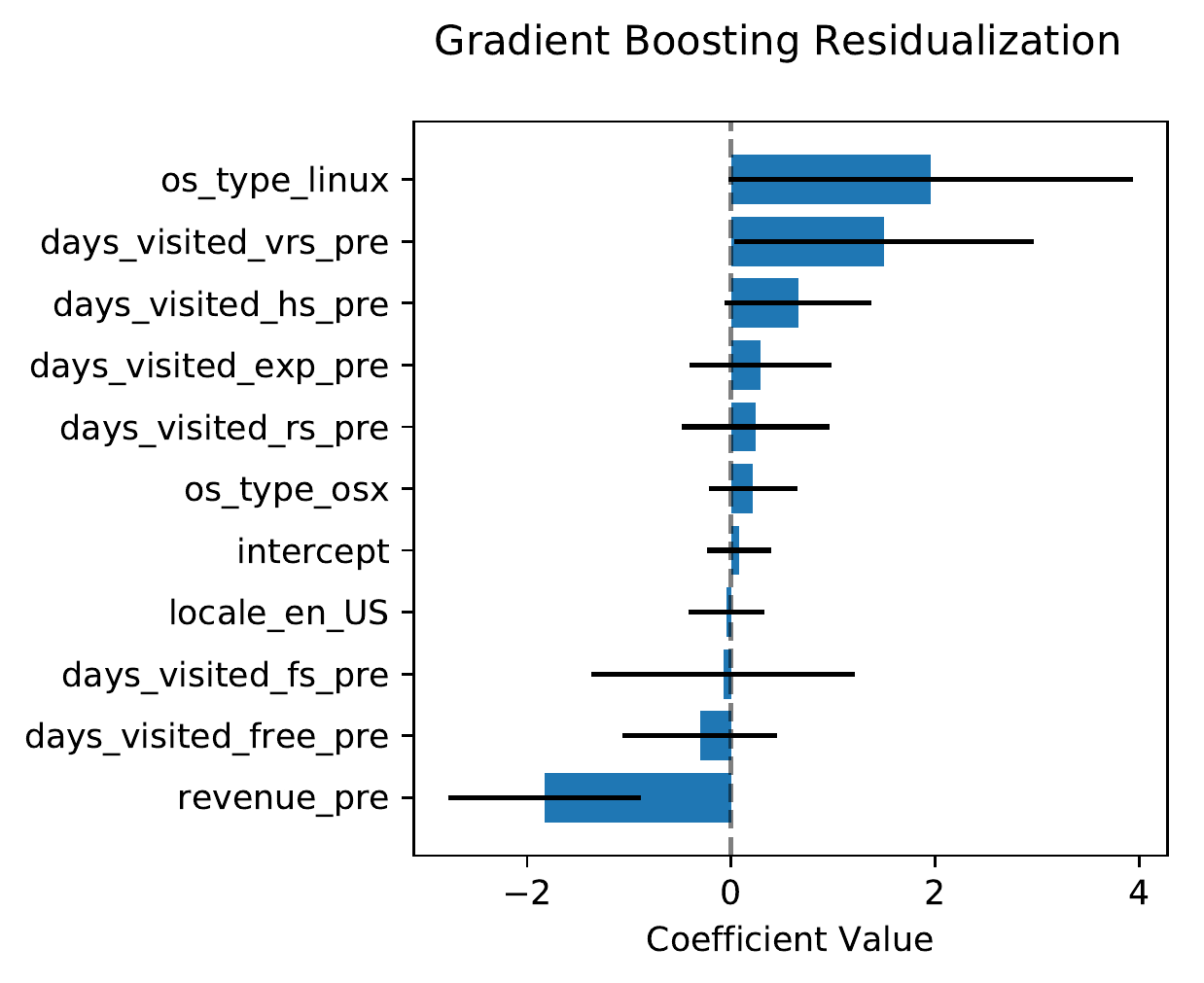}
\end{subfigure}
\caption{(From left to right) Linear CATE projection, SHAP summary of random forest CATE projection, Linear CATE projection coefficients. Using gradient boosting nuisance models.}
\label{fig:3sub-GB-TA}
\end{figure}

Using gradient boosting nuisance models, we show that many inferences remain similar (Figure \ref{fig:3sub-GB-TA} in Appendix). The most notable changes in heterogeneity were for features which have a highly skewed distribution (e.g. visits to specific pages on W), or which appear rarely in the data (e.g. Linux users). 
\vsdelete{This suggests the more powerful models are better able to fit the data for these smaller sub-populations, leading to different results. This explanation is supported by considering the differences between using a linear regression versus random forest for the CATE projection model across residualization models.}
The linear CATE projection model coefficients are largely similar for both residualization models (except the Linux operating system feature, which appears rarely in the data). Moving to a random forest for the CATE projection model with SHAP presents greater differences, especially for the highly skewed features.

\vsdelete{While the point estimates of the ATE under DRIV and DMLATEIV are similar using gradient boosting residualization models, the confidence interval from DRIV is wider than under DMLATEIV. This highlights the potential advantages of DRIV over DMLATEIV in avoiding stronger conclusions about the ATE than could be warranted by the data.}

\textbf{Similar instrument from a recent experiment} A recent 2019 
A/B test of the same membership sign-up process provided another viable instrument. This 21-day A/B test included a much larger, more diverse population of users than in 2018 due to fewer restrictions for eligibility (see Sec. \ref{sec:Wapp-additional-data} for details).
We apply DRIV with gradient boosting residualization models and a linear projection of the CATE. The CATE distribution has generally higher values compared to the 2018 experiment which reflects the different experimental population. In particular, users in the 2018 experiment had much higher engagement and significantly higher revenue in the pre-experiment period. This was largely because users were only included in the 2018 experiment on their \textit{second} visit. The higher baseline naturally makes it more difficult to achieve high treatment effects, explaining the generally lower CATE distribution in the 2018 experiment.
We note that, unlike in 2018, the revenue coefficient is no longer significant. We again attribute this to the much higher revenue baseline in 2018. Despite the population differences, however, \emph{we observe "days\_visited\_vrs\_pre" continues to have a very significant positive association}. "days\_visited\_exp\_pre" now also appears to have a significantly positive association, as does the iPhone device (which was not a feature in the 2018 experiment). \vsedit{The inclusion of iPhone users is another big domain shift in the two experiments.}

\textbf{Policy recommendations for W} Our results offer several policy implications for W. 
Firstly, encourage iPhone users, and users who frequent vacation rentals pages to sign-up for membership. These users exhibited high treatment effects from membership. For frequent visitors to vacation rentals pages, this effect was robust across residualization models, CATE projections, and even different instruments (e.g. by providing stronger encouragements for sign-up on particular sub-pages). 
\vsdelete{Secondly, encourage users who have yet to make a booking (low revenue\_pre) and who are perhaps in the planning and research stage) to sign-up for membership so they can make use of trip-planning tools and personalized travel advice.} 
Second, find ways to improve the membership offering for users who are already engaged: e.g. recently made a booking (high revenue\_pre), were already frequent visitors (high days\_visited\_free\_pre).

\textbf{Validation on Semi-Synthetic Data} In Appendix \ref{sec:W-semi}, we validate the correctness of ATE and CATE from DRIV, by creating a semi-synthetic dataset with the same variables and such that the marginal distribution of each variable looks similar to the TripAdvisor data, but where we know the true effect model. 
We find that DRIV recovers a good estimate of the ATE.
The CATE of DRIV with linear regression as final stage also recovers the true coefficients, and a random forest final stage picks the correct factors of heterogeneity as most important features.
Moreover, coverage of DRIV ATE confidence intervals is almost nominal at 94\%, while DMLATEIV can be very biased and has 26\% coverage.\footnote{Results on the coverage experiment can be recovered by running: \href{https://github.com/microsoft/EconML/tree/master/prototypes/dml_iv/coverage.py}{https://github.com/microsoft/EconML/tree/master/prototypes/dml\_iv/coverage.py} followed by post-processing by \href{https://github.com/microsoft/EconML/tree/master/prototypes/dml_iv/post_processing.ipynb}{https://github.com/microsoft/EconML/tree/master/prototypes/dml\_iv/post\_processing.ipynb}. Single synthetic instance results on the quality of the recovered estimates and how they compare with benchmark approaches can be found in \href{https://github.com/microsoft/EconML/tree/master/prototypes/dml_iv/TA_DGP_Analysis.ipynb}{https://github.com/microsoft/EconML/tree/master/prototypes/dml\_iv/TA\_DGP\_Analysis.ipynb}.}

\section{Estimating the Effect of Schooling on Wages}

The causal impact of schooling on wages has been studied at length in Economics (see \cite{griliches1977estimating}, \cite{card1993using}, \cite{card2001estimating}, \cite{hudson2011parental}), and although it is generally agreed that there is a positive impact, it is difficult to obtain a consistent estimate of the effect due to self-selection into education levels. To account for this endogeneity, Card (\cite{card1993using}) proposes using proximity to a 4-year college as an IV for schooling. We analyze Card's data from the Nat. Long. Survey of Young Men (NLSYM, 1966) to estimate the ATE of education on wages and find sources of heterogeneity. \vsdelete{To validate our findings, we use the NLSYM data to create a semi-synthetic dataset with known treatment effect and show that we correctly recover the true ATE and underlying heterogeneity.}
We describe the NLSYM data in depth in Appendix \ref{sec:nlsyc}. At high level, the data contains 3,010 rows with 22 mostly binary covariates $X$, log wages ($y$), years of schooling ($T$), and 4-year college proximity indicator ($Z$). 

We apply  DMLATEIV and DRIV with linear (LM) or gradient boosted (GBM) nuisance models to estimate the ATE (Table \ref{table:nlsym_ate}). While the DMLATEIV results are consistent with Card's ($0.134, [0.026, 0.242] \;95\%$ CI), this estimate is likely biased in the presence of compliance and effect heterogeneity (see Sec. \ref{sec:warm}). The DRIV ATE estimates, albeit lower, still lie within the 95\% CI of the DML ATE. 

\begin{figure}[H]
\centering
    \begin{subfigure}[b]{0.25\textwidth}
        \includegraphics[height=3.5cm, width=\linewidth]{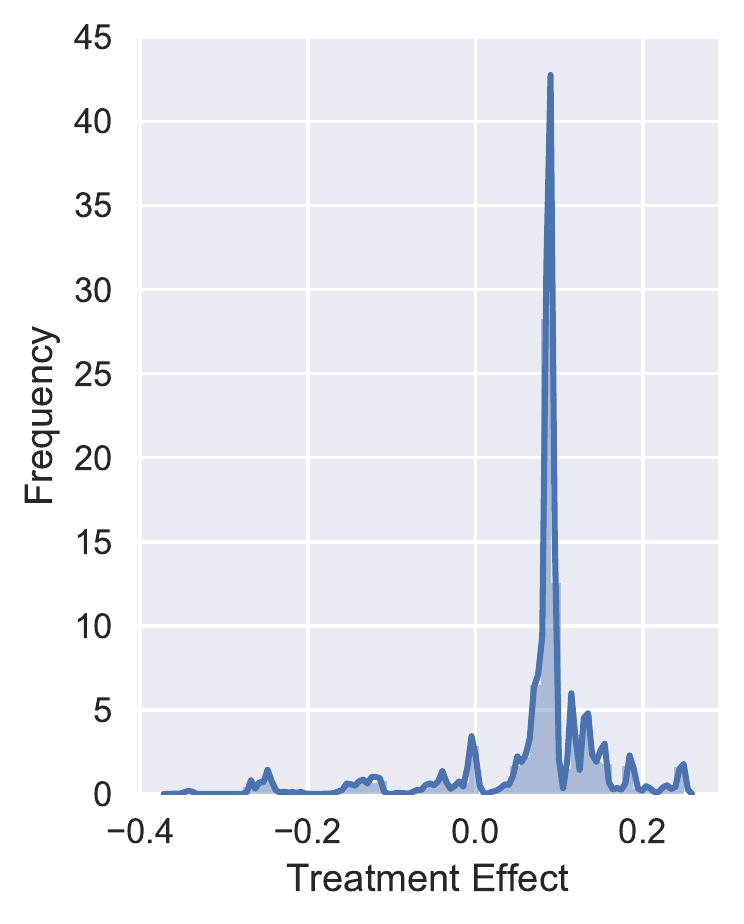} 
        \label{fig:NLSYM_plot1}
    \end{subfigure}
    \begin{subfigure}[b]{0.35\textwidth}
        \includegraphics[height=3.5cm, width=\linewidth]{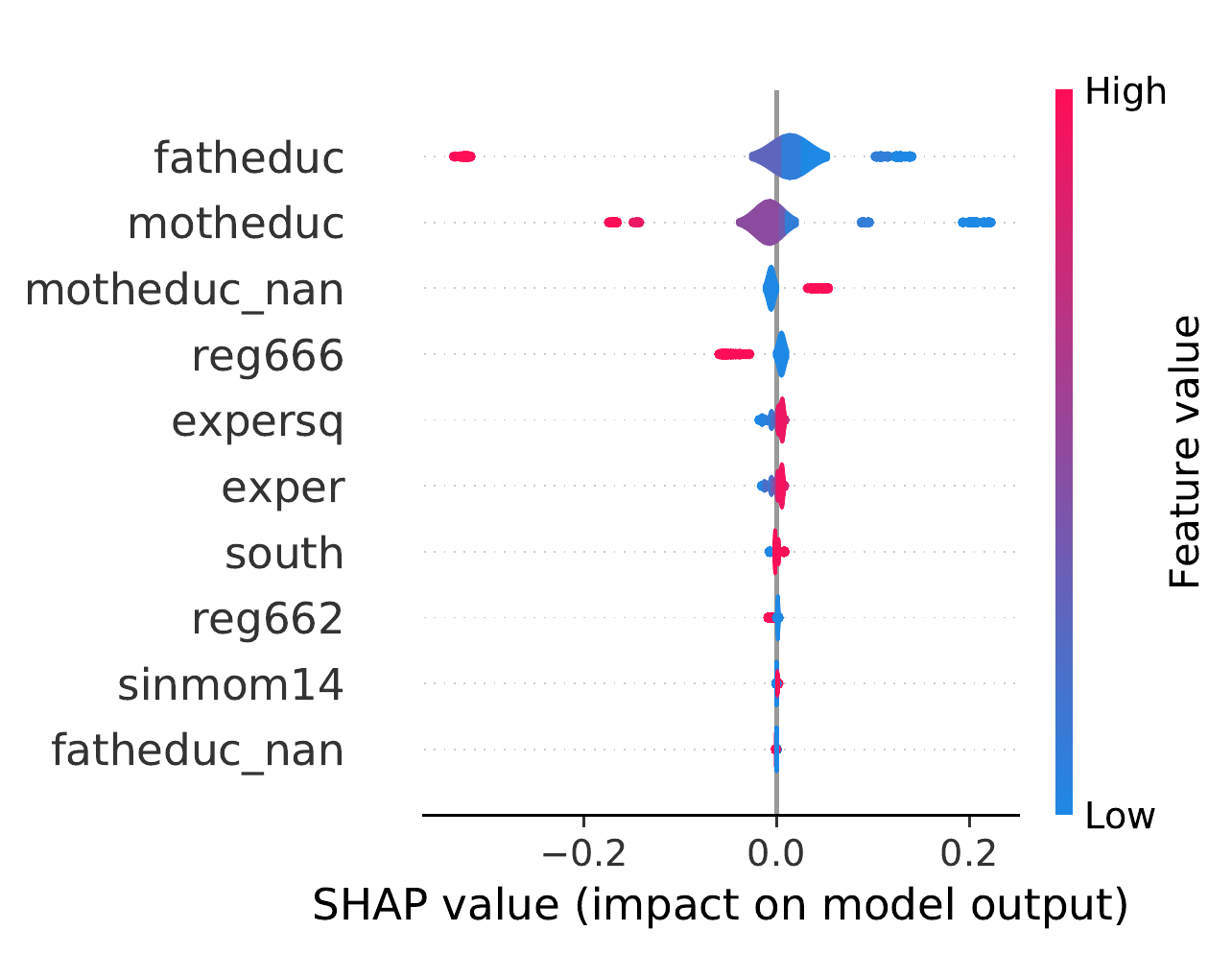} 
        \label{fig:NLSYM_plot2}
    \end{subfigure}
    \begin{subfigure}[b]{0.25\textwidth}
        \includegraphics[height=3.5cm, width=\linewidth]{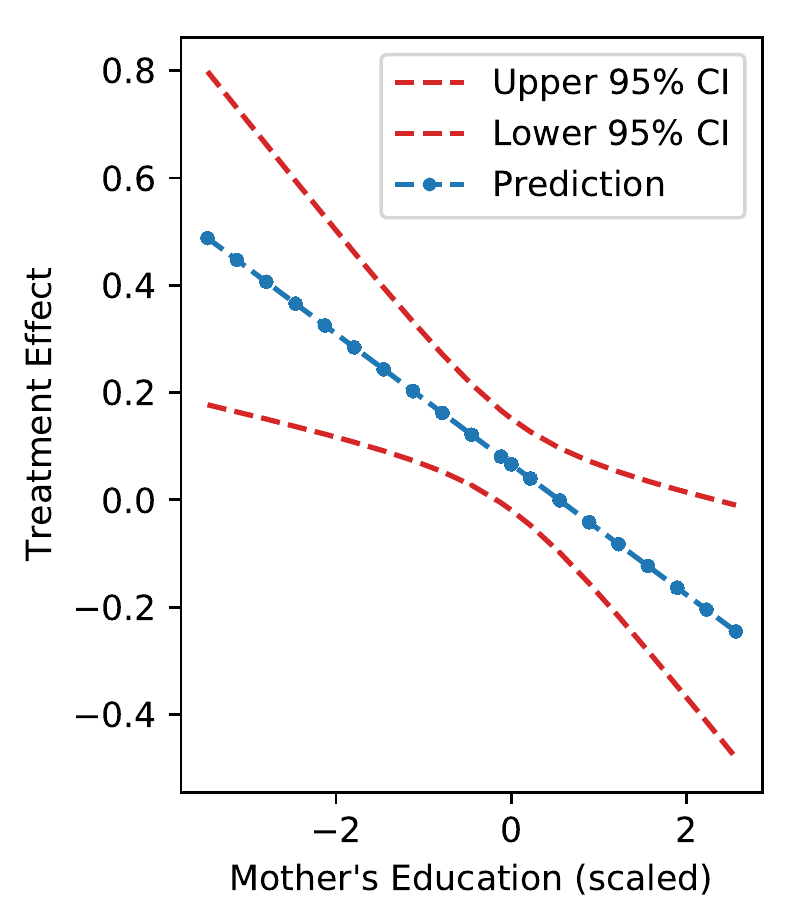}
        \label{fig:NLSYM_plot3}
    \end{subfigure}
\caption{\small{Treatment effect distribution, heterogeneity features, and linear projection on mother's education.}}
\label{fig:image_NLSYM}
\end{figure}

We study effect heterogeneity with a shallow random forest an the last stage of DRIV. Fig. \ref{fig:image_NLSYM} depicts the spread of treatment effects, and the important features selected. Most effects (89\%) are positive, with very few very negative outliers. The heterogeneity is driven mainly by parental education variables. We project the DRIV treatment effect on the mother's education variable to study this effect. In fig. \ref{fig:image_NLSYM}, we note that treatment effects are highest among children of less educated mothers. This pattern has also been observed in \cite{card1993using} and \cite{hudson2011parental}.\footnote{See \href{https://github.com/microsoft/EconML/tree/master/prototypes/dml_iv/NLSYM_Linear.ipynb}{https://github.com/microsoft/EconML/tree/master/prototypes/dml\_iv/NLSYM\_Linear.ipynb} for LM nuisance models and \href{https://github.com/microsoft/EconML/tree/master/prototypes/dml_iv/NLSYM_GBM.ipynb}{https://github.com/microsoft/EconML/tree/master/prototypes/dml\_iv/NLSYM\_GBM.ipynb} for GBM nuisance models for these results.}
\vsdelete{Card (\cite{card1993using}) posits that children with less educated parents are more likely to stop their schooling too soon due to the high costs of education. Considering that the marginal impact of an additional year of schooling is higher in earlier education, the effect will be larger for men in this group. }

\begin{table}[H]
\footnotesize
\centering
\renewcommand\arraystretch{1.1}
\begin{tabular}{|c c|c c|c c c|}\cline{3-7}
\multicolumn{2}{c|}{} & \multicolumn{2}{c|}{Observational Data} & \multicolumn{3}{c|}{Semi-Synthetic Data}\\
 \hline
 Nuisance & Method & ATE Est & 95\% CI & ATE Est & 95\% CI & Cover $\ddag$\\
 \hline
LM & DMLATEIV & 0.137 & [0.027, 0.248] & 0.651 & [0.607, 0.696]$\dagger$ & 52\%\\
LM & DRIV & 0.072 & [0.009, 0.135] & 0.546 & [0.427, 0.665]$\dagger$ & 98\%\\
GBM & DMLATEIV & 0.157 & [0.041, 0.274] & 0.653 & [0.612, 0.694] & 72\%\\
GBM & DRIV & 0.041 & [-0.037, 0.120] & 0.650 & [0.574, 0.727]$\dagger$ & 93\%\\
\hline
\multicolumn{3}{l}{$\dagger$ Contains the true ATE (0.609)} &
\multicolumn{4}{r}{$\ddag$ Coverage for 95\% CI over 100 Monte Carlo simulations}\\
\end{tabular}\\
\caption{\small{NLSYM ATE Estimates for Observational and Semi-synthetic Data}}
\label{table:nlsym_ate}
\end{table}

\textbf{Semi-synthetic Data Results.} We created semi-synthetic data from the NLSYM covariates $X$ and instrument $Z$, with generated treatments and outcomes based on known compliance and treatment functions (see Appx. \ref{sec:nlsyc} for details). 
\vsdelete{We create a realistic treatment effect that depends on the mother's education ($X[4]$) and single mother at age 14 ($X[7]$): $\theta(X) = \; 0.1 + 0.05 \cdot X[4] - 0.1\cdot X[7]$. The DGP is described in full in Appendix \ref{sec:nlsyc}. }
In Table \ref{table:nlsym_ate}, we see that DMLATEIV ATE (true ATE=0.609) is upwards biased and has poor coverage over 100 runs, whereas the DRIV ATE is less biased and has overall good coverage. With DRIV, we also cover the correct $\theta(X)$ with the coefficient CIs when the final stage is the space of linear models on the relevant variables; with linear nuisance models (including products of features): $0.556 \; ([0.431, 0.682] \; 95\% \; \text{CI})$ vs $0.617$, $0.131 \; ([0.010, 0.252])$ vs $0.15$, and $-0.115 \; ([-0.515, 0.285])$ vs $-0.1$; and with gradient boosted forest nuisance models: $0.635 \; ([0.554, 0.716] \; 95\% \; \text{CI})$ vs $0.617$, $0.092 \; ([0.014, 0.169])$ vs $0.15$, and $0.14 \; ([-0.117, 0.397])$ vs $-0.1$. (these are the coefficients after variables where pre-processed and normalized to mean zero and variance $1$) \footnote{See \href{https://github.com/microsoft/EconML/tree/master/prototypes/dml_iv/NLSYM_Semi_Synthetic_Linear.ipynb}{Notebook for Semi-Synthetic Analysis with LM nuisance models} and \href{https://github.com/microsoft/EconML/tree/master/prototypes/dml_iv/NLSYM_Semi_Synthetic_GBM.ipynb}{Notebook for Semi-Synthetic Analysis with GBM nuisance models} for these results.}
\vsdelete{The large CI in the last parameter is due to $X[7]$ being an order of magnitude smaller than $X[4]$.}

\subsubsection*{Acknowledgements}
We thank Jeff Palmucci, Brett Malone, Baskar Mohan, Molly Steinkrauss, Gwyn Fisher and Matthew Dacey from TripAdvisor for their support and assistance in making this collaboration possible.

\bibliographystyle{plain}
\bibliography{references,refs}

\appendix

\section{Proof of Main Lemmas and Corollaries}
Before we prove our two main lemmas we define the concept of an orthogonal loss. Consider a loss function $L(\theta;g)$ that depends on a target model $\theta\in \Theta$ and nuisance model $g\in G$.

\begin{definition}[Directional Derivative]
Let $V$ be a vector space of functions. For a functional $F:V\rightarrow \R$, we define the derivative operator
$$
D_{g}F(g)[\nu] = \frac{d}{dt}F(g+t\nu)\mid{}_{t=0},
$$
for a pair of functions $g,\nu\in V$. Likewise, we define
$$
D_{g}^{k}F(g)[\nu_1,\ldots,\nu_k] = \frac{\partial^{k}}{\partial{}t_1\ldots\partial{}t_k}
F(g + t_1{}\nu_1+\ldots+t_k{}\nu_k)\mid{}_{t_1=\cdots=t_k=0}.
$$
\end{definition}
When considering a functional in two arguments, e.g. $F(\theta,g)$, we will write $D_{g}F(\theta,g)$ and $D_{\theta}F(\theta,g)$ to make the argument with respect to which the derivative is taken explicit.

\begin{definition}[Orthogonal Loss]
\label{ass:orthogonal}
The population risk $L(\theta;g)$ is \emph{orthogonal}, if:
\begin{equation}
\label{eq:orthogonal}
D_{g}D_{\theta} L(\theta_0; g_0)[\theta-\theta_0,g-g_0] = 0\quad\forall{}\theta\in\Theta,\forall{} g\in G.
\end{equation}
\end{definition}

Suppose that the loss function is the expectation of a point-wise loss:
\begin{equation}
    L(\theta; g) = \E[\ell(U; \theta(X), g(W))]
\end{equation}
where $U$ represents all random variables of the data generating process and $X, W$ are subsets of these variables. 

Then orthogonality is implied by the condition, $\forall \theta\in \Theta$, $\forall g\in G$:
\begin{equation}
    \E[\left(g(W)-g_0(W)\right)^T \nabla_{g(W)}\, m(U; \theta_0(X), g_0(W))\, \left(\theta(X)-\theta_0(X)\right)]=0
\end{equation}
where $m(U; \theta(X), g(W))=\nabla_{\theta(X)}\ell(U;\theta(X), g(W))$. We will typically refer to $m$ as the moment that corresponds to the loss function $\ell$. Subsequently, it suffices that:
\begin{equation}
    \E[\left(g(W)-g_0(W)\right)^T \nabla_{g(W)}\, m(U; \theta_0(X), g_0(W)) \mid X]=0
\end{equation}
or the even stronger condition that:
\begin{equation}
    \E[\nabla_{g(W)}\, m(U; \theta_0(X), g_0(W)) \mid X, W]=0
\end{equation}
In most Lemmas in the next few sections we will show the latter stronger sufficient condition.

\subsection{Proof of Lemma~\ref{thm:orthogonality}}
\begin{proof}
We show that the expected directional derivative of the moment (directional derivative of the loss with respect to $\theta(X)$) conditional on $X$, with respect to each of the nuisance functions is equal to zero, when evaluated at the true nuisance and target functions.
The directional derivative of the loss with respect to direction $\nu=\theta'-\theta$ and evaluated at parameter $\theta$ is:
\begin{align*}
    \E[m^1(X; \theta(X), q(X), p(X), h)\cdot \nu(X)]
\end{align*}
where:
\begin{align*}
    m^1(X; \theta(X), q(X), p(X), h) = - 2\, \E[ \left(Y - q(X) - \theta(X)\, \left(h(Z, X) - p(X)\right)\right) \, \left(h(Z, X) - p(X)\right)\mid X]
\end{align*}
To show orthogonality with respect to $p, q$, it suffices to show that the classical derivative of $m^1$ with respect to the inputs $p(X)$ and $q(X)$ is zero, when evaluated at the true nuisance and target parameters:
\begin{align*}
    \nabla_{q(X)} m^1(X; \theta_0(X), q_0(X), p_0(X), h_0) :=~& -2\,\E[h_0(Z, X) - p_0(X) \mid X]=0\\
    \nabla_{p(X)} m^1(X; \theta_0(X), q_0(X), p_0(X), h_0) :=~& -2\, \theta_0(X)\,\E[h_0(Z, X) - p_0(X) \mid X]\\
    ~& + 2\, \E[Y - q_0(X) - \theta_0(X)\, \left(h_0(Z, X) - p_0(X)\right)\mid X]\\
    =~& 0
\end{align*}
Where in both equations we invoked the conditional moment restrictions to claim that they are equal to zero.

To prove orthogonality with respect to $h$ we need to show that the directional derivative of $m^1$ with respect to $h$ is zero. We cannot reduce it to a classical derivative condition, since $h$ takes as input the variable $Z$ which is not part of the conditioning set of the moment $m^1$. However, we see that this directional derivative evaluated at $h_0$ and at a direction $\nu=h-h_0$, is not zero:
\begin{align*}
    \D_h m^1(X; \theta_0, q_0, p_0, h_0)[\nu] :=~& 2\,\theta_0(X)\,\E[(h_0(Z, X) - p_0(X))\, \nu(Z, X) \mid X]\\
    ~&  + 2\, \E[(Y - q_0(X) - \theta_0(X)\, \left(h_0(Z, X) - p_0(X)\right))\, \nu(Z, X)\mid X]\\
    =~& 2\,\theta_0(X)\,\E[(h_0(Z, X) - p_0(X))\, \nu(Z, X) \mid X]\\
\end{align*}
The last quantity is not necessarily zero, since $\E[h_0(Z, X)- p_0(X)\mid Z, X]\neq 0$. This finding is reasonable since we are using $h(Z, X)$ as our regressor. Hence, any error in the measurement of the regressor should directly propagate to an error in $\theta(X)$. The quantity would have been zero if the residual error from the first stage function $h(Z, X)-h_0(Z,X)$ was independent of the residual randomness $h_0(Z, X)-p_0(X)$, conditional on $X$. However, the two in general can be correlated: the second quantity measures how far is $h_0(Z, X)$ from each mean $p_0(X)=\E[h_0(Z, X)\mid X]$, while the first quantity measures how far is the estimate $h(Z, X)$ from $h_0(Z, X)$. It is highly probable that when $Z$ takes values that lead to a large deviation from the mean treatment, then these are also the values of $Z$ for which the first stage model makes more mistakes.
\end{proof}

\subsection{Proof of Lemma~\ref{thm:dr-orthogonality}}
\begin{proof}
We show that the expected derivative of the moment (derivative of the loss with respect to $\theta(X)$) conditional on $X$, with respect to each of the nuisance functions is equal to zero, when evaluated at the true nuisance and target functions. The directional derivative of the loss with respect to direction $\nu=\theta'-\theta$ and evaluated at parameter $\theta$ is:
\begin{align*}
    -2\,\E[m^2(X; \theta(X), g(X))\cdot \nu(X)]
\end{align*}
where $g(X) = (\theta_{pre}(X), p(X), q(X), r(X), \beta(X))$ and :
\begin{align*}
    m^2(X; \theta(X), g(X)) = \E\left[ \theta_{pre}(X) + \frac{(Y - q(X) - \theta_{pre}(X)\, (T - p(X)))\, (Z-r(X))}{\beta(X)} - \theta(X)\bigg| X\right]
\end{align*}
To show orthogonality with respect to the nuisance functions $g$, it suffices to show that the classical derivative of $m^2$ with respect to each component of $g(X)$ is zero, when evaluated at the true nuisance and target parameters:
\begin{align*}
    \nabla_{\theta_{pre}(X)} m^2(X; \theta_0(X), g_0(X)) :=~& \E\left[1 - \frac{ (T - p_0(X))\, (Z-r_0(X))}{\beta_0(X)}\mid X\right]=0\\
    =~& 1 - \frac{\E[(T - p_0(X))\, (Z-r_0(X))\mid X]}{\beta_0(X)} = 0\\
    \nabla_{p(X)} m^2(X; \theta_0(X), g_0(X)) :=~&  \theta_0(X)\,\frac{\E[Z - p_0(X) \mid X]}{\beta_0(X)}=0\\
    \nabla_{q(X)} m^2(X; \theta_0(X), g_0(X)) :=~&  -\frac{\E[Z - r_0(X) \mid X]}{\beta_0(X)}=0\\
    \nabla_{r(X)} m^2(X; \theta_0(X), g_0(X)) :=~&  - \frac{\E[Y-q_0(X)\mid X]}{\beta_0(X)} + \theta_0(X)\,\frac{\E[T - p_0(X) \mid X]}{\beta_0(X)}=0\\
    \nabla_{\beta(X)} m^2(X; \theta_0(X), g_0(X)) :=~& -\frac{\E\left[(Y - q_0(X) - \theta_0(X)\, (T - p_0(X)))\, (Z-r_0(X))\mid X\right]}{\beta_0(X)^2}\\
    =~& -\frac{\E\left[\E[Y - q_0(X) - \theta_0(X)\, (T - p_0(X)) \mid Z, X]\, (Z-r_0(X))\mid X\right]}{\beta_0(X)^2}\\
    =~& -\frac{\E\left[\E[Y - \theta_0(X)\, T - f_0(X) \mid Z, X]\, (Z-r_0(X))\mid X\right]}{\beta_0(X)^2} = 0
\end{align*}
Where in all equations we invoked the conditional moment restrictions and the definitions of the true nuisance functions to claim that they are equal to zero.
\end{proof}

\subsection{Proof of Lemma~\ref{thm:dr-orthogonality-well-spec}}
\begin{proof}
We show that the expected derivative of the moment (derivative of the loss with respect to $\theta(X)$) conditional on $X$, with respect to each of the nuisance functions is equal to zero, when evaluated at the true nuisance and target functions. The directional derivative of the loss with respect to direction $\nu=\theta'-\theta$ and evaluated at parameter $\theta$ is:
\begin{align*}
    -2\,\E[m_{rw}^2(X; \theta(X), g(X))\cdot \nu(X)]
\end{align*}
where $g(X) = (\theta_{pre}(X), p(X), q(X), r(X), \beta(X))$ and $m_{rw}^2(X; \theta(X), g(X))$ is equal to:
\begin{align*}
    \E\left[ \beta(X)\, \left(\theta_{pre}(X)\beta(X) + (Y - q(X) - \theta_{pre}(X)\, (T - p(X)))\, (Z-r(X)) - \theta(X)\beta(X)\right)\bigg| X\right]
\end{align*}
To show orthogonality with respect to the nuisance functions $g$, it suffices to show that the classical derivative of $m_{rw}^2$ with respect to each component of $g(X)$ is zero, when evaluated at the true nuisance and target parameters:
\begin{align*}
    \nabla_{\theta_{pre}(X)} m_{rw}^2(X; \theta_0(X), g_0(X)) :=~& \E\left[\beta_0(X)\, (\beta_0(X) - (T - p_0(X))\, (Z-r_0(X)))\mid X\right]=0\\
    =~& \beta_0(X)\, \left(\beta_0(X) - \E[(T - p_0(X))\, (Z-r_0(X))\mid X]\right) = 0\\
    \nabla_{p(X)} m_{rw}^2(X; \theta_0(X), g_0(X)) :=~&  \beta_0(X)\, \theta_0(X)\,\E[Z - p_0(X) \mid X]=0\\
    \nabla_{q(X)} m_{rw}^2(X; \theta_0(X), g_0(X)) :=~&  -\beta_0(X)\E[Z - r_0(X) \mid X]=0\\
    \nabla_{r(X)} m_{rw}^2(X; \theta_0(X), g_0(X)) :=~&  -\beta_0(X) \left(\E[Y-q_0(X)\mid X] - \theta_0(X)\,\E[T - p_0(X) \mid X]\right)=0\\
    \nabla_{\beta(X)} m_{rw}^2(X; \theta_0(X), g_0(X)) :=~& 
    \E\left[(Y - q_0(X) - \theta_0(X)\, (T - p_0(X)))\, (Z-r_0(X))\mid X\right]=0
\end{align*}
Where in all equations we invoked the conditional moment restrictions and the definitions of the true nuisance functions to claim that they are equal to zero.
\end{proof}

\subsection{Proof of Lemma~\ref{thm:projected-dr-orthogonality}}
\begin{proof}
We show that the expected derivative of the moment (derivative of the loss with respect to $\theta(X)$) conditional on $X$, with respect to each of the nuisance functions is equal to zero, when evaluated at the true nuisance and target functions. The directional derivative of the loss with respect to direction $\nu=\theta'-\theta$ and evaluated at parameter $\theta$ is:
\begin{align*}
    -2\,\E[m_{\pi,rw}^2(X; \theta(X), g(X))\cdot \nu(X)]
\end{align*}
where $g(X) = (\theta_{pre}(X), p(X), q(X), h(X), V(X))$ and  $m_{rw}^2(X; \theta(X), g(X))$ is equal to:
\begin{align*}
\E\left[ V(X)\, \left(\theta_{pre}(X)V(X) + (Y - q(X) - \theta_{pre}(X)\, (T - p(X)))\, (h(Z, X)-p(X)) - \theta(X)V(X)\right)\bigg| X\right]
\end{align*}
To show orthogonality with respect to the nuisance functions $g$, it suffices to show that the classical derivative of $m_{\pi, rw}^2$ with respect to each component of $g(X)$ is zero, when evaluated at the true nuisance and target parameters:
\begin{align*}
    \nabla_{\theta_{pre}(X)} m_{\pi, rw}^2(X; \theta_0(X), g_0(X)) :=~& \E\left[V_0(X)\, (V_0(X) - (T - p_0(X))\, (h_0(Z, X)-p_0(X)))\mid X\right]=0\\
    =~& V_0(X)\, \left(V_0(X) - \E[(T - p_0(X))\, (h_0(Z,X)-p_0(X))\mid X]\right) = 0\\
    \nabla_{p(X)} m_{\pi, rw}^2(X; \theta_0(X), g_0(X)) :=~&  V_0(X)\, \theta_0(X)\,\left(\E[h_0(Z, X) - p_0(X) \mid X] + \E[T - p_0(X)\mid X]\right)\\
    &~~ - V_0(X)\E[Y - q_0(X)\mid X]\\
    =~& 0\\
    \nabla_{q(X)} m_{\pi, rw}^2(X; \theta_0(X), g_0(X)) :=~&  -V_0(X)\E[h_0(Z, X) - p_0(X) \mid X]=0\\
    \nabla_{h(Z, X)} m_{\pi, rw}^2(X; \theta_0(X), g_0(X)) :=~&  V_0(X)\left(\E[Y-q_0(X)\mid X] - \theta_0(X)\E[T-p_0(X)\mid X]\right)=0\\
    \nabla_{V(X)} m_{\pi, rw}^2(X; \theta_0(X), g_0(X)) :=~& 
    \E\left[(Y - q_0(X) - \theta_0(X)\, (T - p_0(X)))\, (h_0(Z,X)-p_0(X))\mid X\right]=0
\end{align*}
Where in all equations we invoked the conditional moment restrictions and the definitions of the true nuisance functions to claim that they are equal to zero.
\end{proof}

\subsection{Proof of Main Corollaries}

Observe that all our loss functions fall into the single-index setting presented in Section 4.1 of \cite{foster2019orthogonal}, as all of them are of the form:
\begin{equation}\label{eqn:square-loss}
    L(\theta;g) = \E[(\Gamma(U; g(W)) - \Lambda(U; g(W))\, \theta(X))^2]
\end{equation}
In which case the ex-post loss is 
\begin{equation}
    \ell(U; \theta(X), g(W))=(\Gamma(U; g(W)) - \Lambda(U; g(W))\, \theta(X))^2
\end{equation} 
and the moment is
\begin{equation}
m(U; \theta(X), g(W)) = 2 (\Gamma(U; g(W)) - \Lambda(U; g(W))\, \theta(X))\, \Lambda(U; g(W))    
\end{equation} 
More concretely, the quantities $\Gamma$ and $\Lambda$ for each of our loss functions are:
\begin{enumerate}
    \item For loss $L^1(\theta; q, h, p)$ : 
    \begin{align*}
        \Gamma(U;g(W))=~& Y-q(X)\\
        \Lambda(U;g(W))=~& h(Z, X)-p(X)
    \end{align*}
    \item For loss $L^2(\theta; \theta_{pre}, \beta, p, q, r)$:
    \begin{align*}
        \Gamma(U;g(W))=~&\theta_{pre}(X) + \frac{(Y-q(X) - \theta_{pre}(X)(T-p(X)))\, (Z-r(X))}{\beta(X)}\\
        \Lambda(U;g(W))=~& 1
    \end{align*}
    \item For loss $L_{rw}^2(\theta; \theta_{pre}, \beta, p, q, r)$: 
    \begin{align*}
        \Gamma(U; g(W))=~& \beta(X)\theta_{pre}(X) + (Y-q(X) - \theta_{pre}(X)(T-p(X)))\, (Z-r(X))\\
        \Lambda(U;g(W))=~&\beta(X)
    \end{align*}
    \item For loss $L_{\pi, rw}^2(\theta; \theta_{pre}, V, p, q, h)$:
    \begin{align*}
        \Gamma(U; g(W))=~& V(X)\theta_{pre}(X) + (Y-q(X) - \theta_{pre}(X)(T-p(X)))\, (h(Z,X)-r(X))\\
        \Lambda(U;g(W))=~& V(X)
    \end{align*}
\end{enumerate}

Thus to prove the main corollaries it suffices to verify that Assumption~7 of \cite{foster2019orthogonal} is satisfied for each of the losses. We first present Assumption~7 in our notation and for the special case of square losses and then verify each of the conditions holds for our losses.
\begin{assumption}[\cite{foster2019orthogonal}] Consider any loss function $L(\theta;g)$ of the form presented in Equation~\eqref{eqn:square-loss} and let $\theta_*$ be a minimizer within space $\Theta$. Then the following must hold:
\begin{enumerate}
    \item The loss $L$ is orthogonal, i.e. $\forall g\in G$: \begin{equation*}
        \E[(g(W)-g_0(W))^T \nabla_{g(W)} m(U; \theta_*(X), g_0(X))\mid X]=0
    \end{equation*}
    \item $\theta_*$ is first order optimal: $\forall \theta\in \Theta$: \begin{equation*}
        \E[m(U; \theta_*(X), g_0(W))\, (\theta(X)-\theta_*(X))]\geq 0
    \end{equation*}
    \item $\Lambda(U; g(W))$ is Lipschitz in $g(W)$, w.r.t. the $\|\cdot\|_2$ norm a.s.
    \item $\sup_{X, \theta\in \Theta}|\theta(X)|$ is upper bounded by a constant
    \item The following strong convexity condition is satisfied:
    \begin{equation*}
    \E\left[\Lambda(U; g_0(W))^2\, \left(\theta(X)-\theta_*(X)\right)^2\right] \geq \gamma\,  \E\left[(\theta(X)-\theta_*(X))^2\right]
    \end{equation*}
    \item The moment is second order smooth with respect to $g(W)$: for all $g\in G$ the spectral norm of the Hessian of the moment w.r.t. $g(W)$, $\|\nabla_{g(W), g(W)}^2 m(U; \theta_*(X), g(W))\mid W]\|_{\sigma}$, is upper bounded by a constant.
\end{enumerate}
\end{assumption}

We now verify that each of these conditions holds:
\begin{enumerate}
    \item The first part is the orthogonality condition which is implied by the Lemmas in the previous sections. The only exception is loss $L^1$, which is not orthogonal with respect to $h$.
    \item The second part is always satisfied whenever $\Theta_{\pi}$ is a convex function space by first order optimality of $\theta_*$. If it is not convex, but there exists $\theta_0\in \Theta_{\pi}$ such that:
    \begin{equation}
        \E[m(U; \theta_0(X), g_0(W)) \mid X]=0
    \end{equation}
    Then this is also satisfied. The former will be the case when we are doing projections over simpler hypothesis spaces, while the latter is the case when $\Theta_{\pi}$ contains the true CATE.
    \item The third part is satisfied since $\Lambda(U; g(W))$ is equal to $h(Z, X)-p(X)$ or $1$ or $\beta(X)$ or $V(X)$ respectively for $L^1, L^2, L_{rw}^2$ and $L_{\pi, rw}^2$. Hence, it is always $1$-Lipschitz in $g(W)$.
    \item The fourth part is satisfied whenever all variables are bounded, since then $\sup_{X, \theta\in \Theta_{\pi}} |\theta(X)|$ is bounded by a constant.
    \item The fifth part is a strong convexity condition that is assumed in each of our corollaries.
    This is satisfied for loss $L^1$ if the following average overlap condition holds:
    \begin{equation}
        \E[V_0(X)\, \left(\theta(X)-\theta_0(X)\right)^2] \geq \gamma\, \E[(\theta(X)-\theta_0(X))^2]
    \end{equation}
    For loss $L^2$ it is satisfied with constant $1$, since $\Lambda(U;g_0(W))=1$. For loss $L_{rw}^2$ it is satisfied if the following average overlap condition holds:
    \begin{equation}
        \E[\beta_0(X)^2\, \left(\theta(X)-\theta_0(X)\right)^2]\geq \gamma\,  \E[(\theta(X)-\theta_0(X))^2]
    \end{equation}
    and similarly for loss $L_{\pi, rw}^2$ if:
    \begin{equation}
        \E[V_0(X)^2\, \left(\theta(X)-\theta_0(X)\right)^2]\geq \gamma\,  \E[(\theta(X)-\theta_0(X))^2]
    \end{equation}
    \item The sixth condition requires that the moment have a bounded second derivative with respect to $g(W)$. Observe that if all the random variables are bounded then losses $L^1$, $L_{rw}^2$ and $L_{\pi, rw}^2$ are smooth and hence all derivatives are bounded. Moreover, loss $L^2$ is smooth if further $\beta(X)>\beta_{\min}>0$ for all $X$.
\end{enumerate} 

Thus all conditions of Assumption~7 of \cite{foster2019orthogonal} are satisfied. Therefore, Corollaries~\ref{cor:driv}, \ref{cor:driv-rw}, \ref{cor:projected-driv-rw} follow directly from Corollary~1 of \cite{foster2019orthogonal}. Corollary~\ref{cor:dmliv} follows by straightforward modification of the proof of Corollary~1, to account for the non-orthogonality with respect to $h$ (hence we omit the adaptation). 

\section{TripAdvisor Data and Analysis}
\subsection{Model Details and Parameters}\label{sec:Wapp-model-details}

\textbf{Residualization Models}
\begin{itemize}
    \item LASSO regression and logistic regression with an L2 penalty using the Python sklearn library. For each cross-fitted fold, 3-Fold cross-validation was used to select the regularization parameter based on minimizing RMSE and log-loss.
    \item Gradient boosting (GB) regression and classification using the XGBoost library.\cite{Chen:2016:XST:2939672.2939785} 100 estimators were used, with a minimum child weight of 20, and gamma set to 0.1. A 10\% validation set was used for early stopping based on RMSE and log-loss.
\end{itemize}

The gradient boosting models from sklearn also yielded substantially similar results to XGBoost.

\textbf{Random Forest} We use a shallow, heavily regularized random forest for projection of the CATE. Parameters used: 1,000 trees, a minimum leaf size of 20,000, and a maximum depth size of 1. The heavy regularization is required in order to ensure stability of the CATE estimates.

\begin{figure}[H]
    \centering
    \includegraphics[width=.75\linewidth]{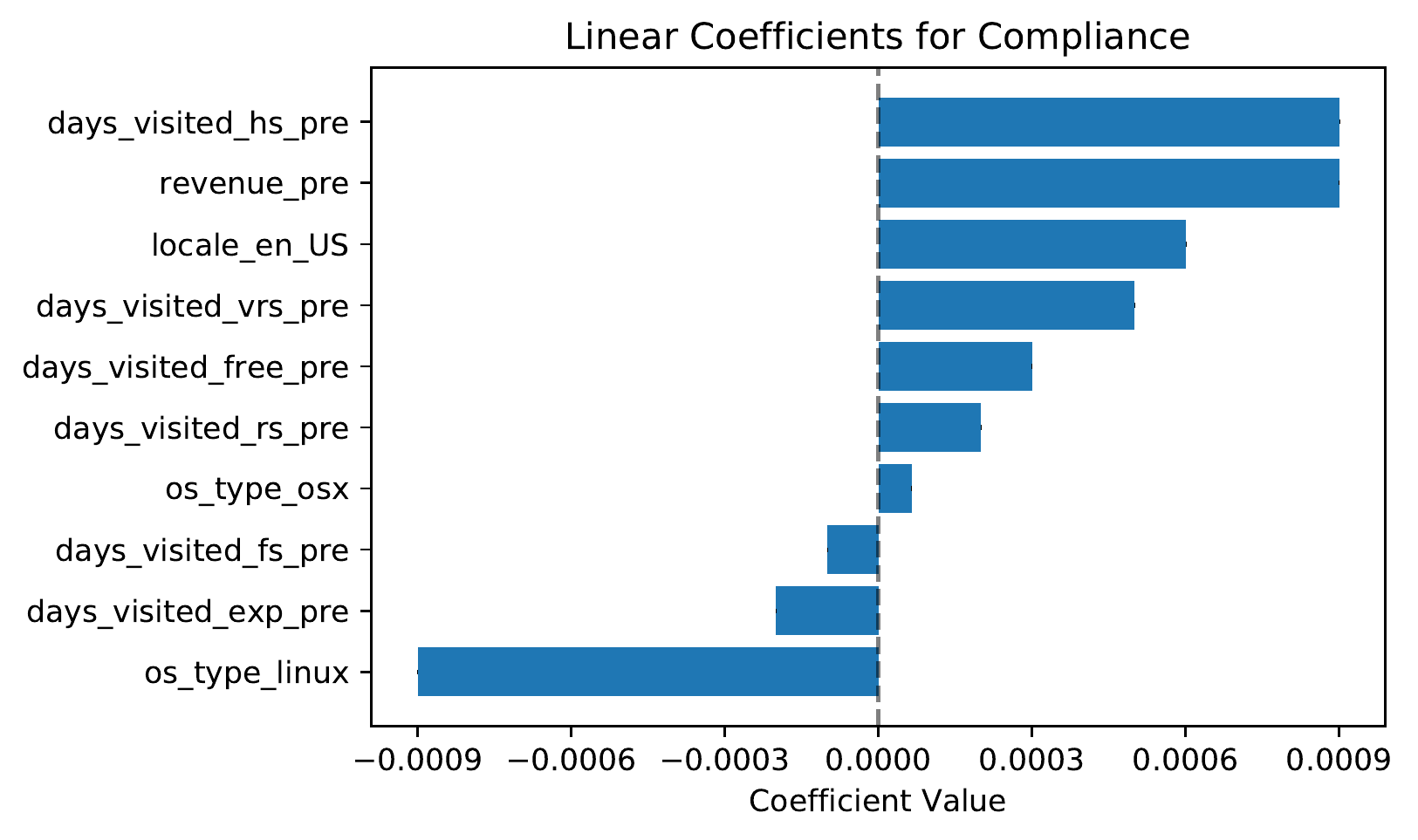}
    \caption{Coefficients of the linear model approximation of the compliance quantity $E[T \cdot Z|X] - E[T|X] \cdot E[Z|X]$. Using linear models for nuisance.}
    \label{fig:linear-compliance-coefficients-linear-resid}
\end{figure}

\textbf{Linear Compliance Model}
Using the 2018 experiment data and linear residualization models, the compliance quantity $E[T \cdot Z|X] - E[T|X] \cdot E[Z|X]$ (despite the logistic function) is well-approximated by a linear regression. We use this approximation for interpreting the coefficents of the fitted model (Figure \ref{fig:linear-compliance-coefficients-linear-resid}).

\subsection{Additional Data Description and Preparation} \label{sec:Wapp-additional-data}

Full description of the data in Table \ref{table:data_desc_ta}. The criteria for eligibility required that users were not existing members of TripAdvisor before the experimental period; visited TripAdvisor through a desktop browser during the experimental period; and visited TripAdvisor at least twice during the experimental period. The first visit did not activate the test functionality. Group assignment was determined randomly with equal probability, resulting in $n_A = 2,303,658$ in group A, and $n_B = 2,302,383$ in group B. 

We transform the operating system categorical variable using one-hot encoding and drop the "Windows" level to use as the baseline. In addition, the co-variates are normalized uniformly over 1,000 quantiles, resulting in a $X_i \in [0, 1]^{10}$ co-variate vector for each user to be used for both residualization and effect heterogeneity.

For confidentiality reasons, we report the ATE and CATE results normalized by $\hat{\mu}_B$, the mean number of days visited by users in group B of the A/B experiment. A treatment effect of 1 unit is therefore equal to $\hat{\mu}_B$ additional days visited.

\begin{table}[H]
\renewcommand\arraystretch{1.5}
\begin{tabular}{p{0.23\linewidth}p{0.70\linewidth}} 
 \hline
revenue\_pre & Total revenue in dollars generated by the user in the pre-experimental period \\ 
 days\_visited\_free\_pre & Count of the days the user visited the TripAdvisor through free channels (e.g. email) in the pre-experimental period (0-28) \\
 days\_visited\_hs\_pre & Count of the days the user visited the hotels pages of TripAdvisor in the pre-experimental period (0-28) \\
 days\_visited\_exp\_pre & Count of the days the user visited the experiences pages of TripAdvisor in the pre-experimental period (0-28) \\
 days\_visited\_rs\_pre & Count of the days the user visited the restaurants pages of TripAdvisor in the pre-experimental period (0-28) \\
 days\_visited\_vrs\_pre & Count of the days the user visited the vacation rentals pages of TripAdvisor in the pre-experimental period (0-28) \\
 days\_visited\_fs\_pre & Count of the days the user visited the flights pages of TripAdvisor in the pre-experimental period (0-28) \\
 os\_type & Categorical variable for the user's operating system (3 levels) \\
 locale\_en\_US & Binary variable indicating whether the user was from the en\_US locale \\
 \hline
 Y & Outcome measurement, count of the number of total days the user visited TripAdvisor \\
 T & Treatment, binary variable of whether the user became a member during the experimental period \\
 Z & Instrument, binary variable of the user's group assignment in the A/B test \\
 \hline
\end{tabular}
\caption{Definition of variables in the 2018 experimental data from W}
\label{table:data_desc_ta}
\end{table}

\textbf{Additional details about the 2019 experiment} There were some key differences compared to the 2018 A/B test:

\begin{itemize}
    \item the test was run for 3 weeks instead of 2;
    \item the test functionality was displayed on both desktop and mobile platforms across nearly all pages of TripAdvisor (i.e. not just the homepage);
    \item first-time visitors were eligible for the test; and
    \item the sample size was much larger at $n = 84,657,263$ users. We use a sample of $n_S = 10,158,871$ users stratified by A/B test group allocation for computational reasons.
\end{itemize}

\subsection{Additional Figures of Experimental Results}

\begin{figure}[H]
    \centering
    \includegraphics[width=.75\linewidth]{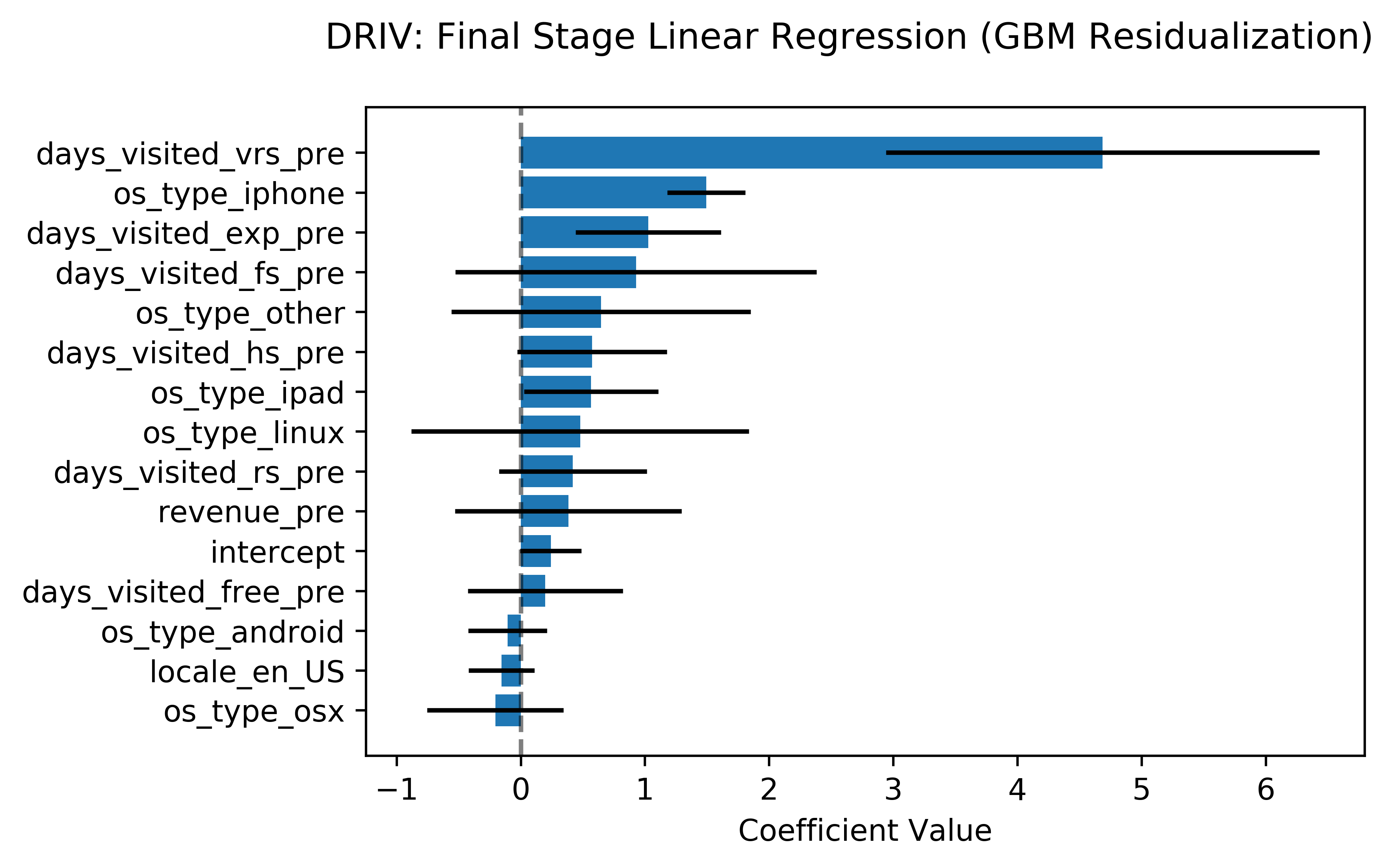}
    \caption{Coefficients of the CATE linear projection model using DRIV with gradient boosting residualization on the TripAdvisor 2019 experiment data.}
    \label{fig:10M_xgb_final_linear}
\end{figure}

\begin{figure}[H]
\begin{subfigure}{.5\textwidth}
    \centering
    \includegraphics[width=.95\linewidth]{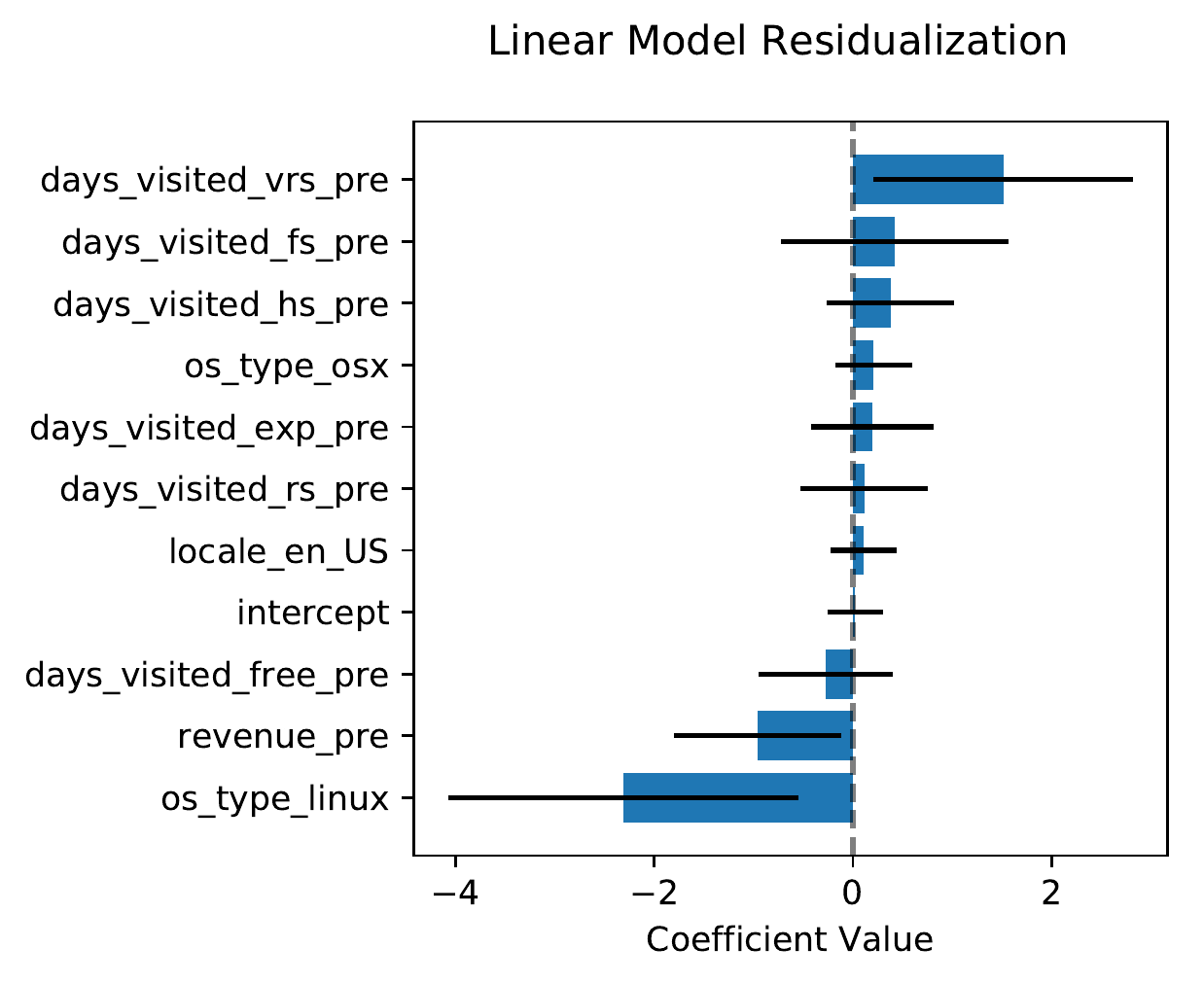}
    
\end{subfigure}%
\begin{subfigure}{.5\textwidth}
    \centering
    \includegraphics[width=.95\linewidth]{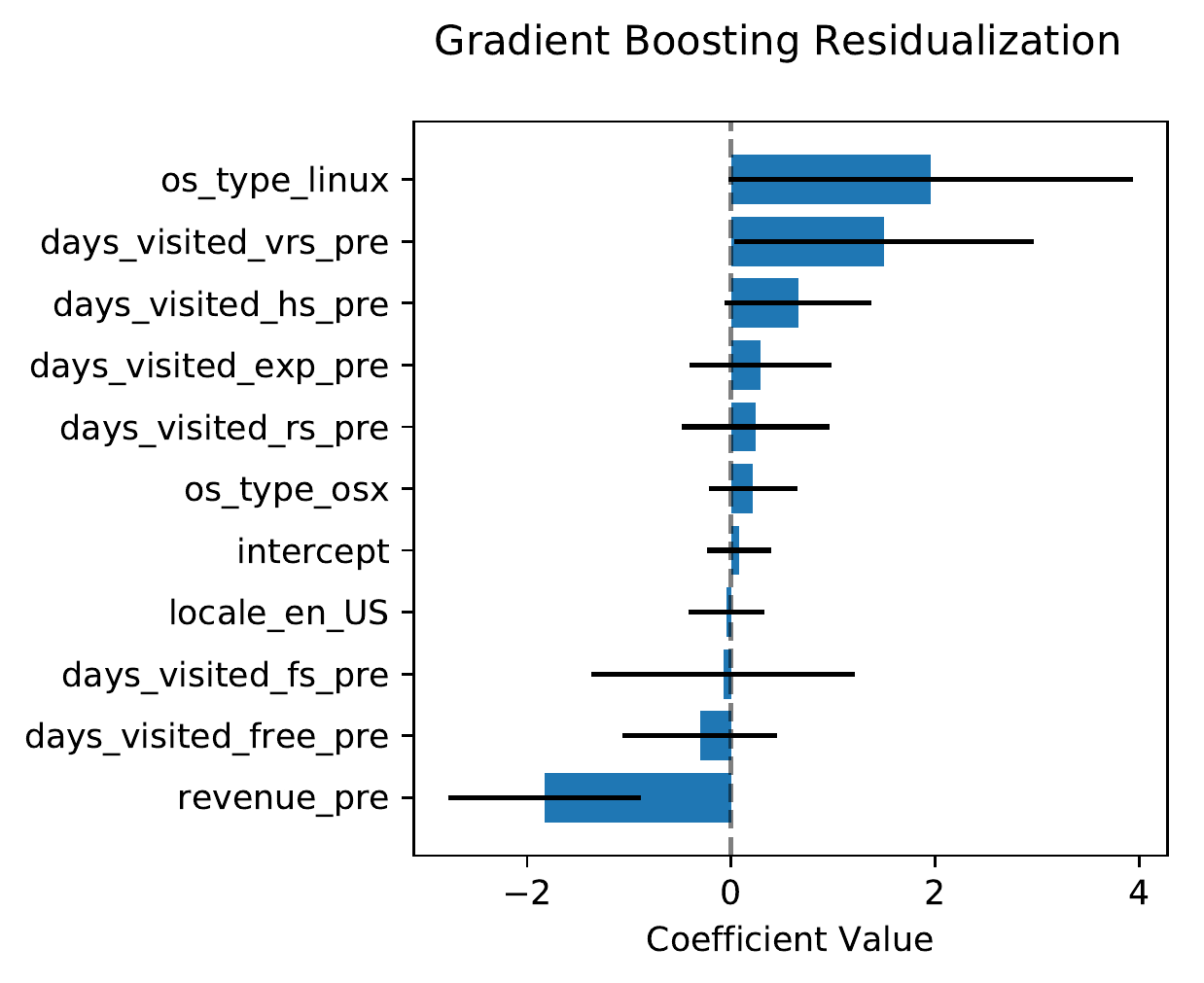}
\end{subfigure}
\caption{Coefficients of the linear CATE projection model for DRIV}
\label{fig:linear-coef-linear-resid}
\end{figure}

\begin{figure}[H]
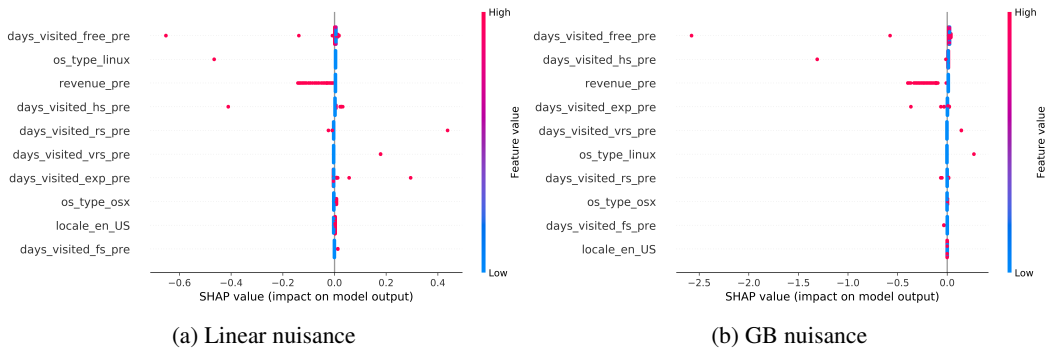

\begin{subfigure}{.5\textwidth}
    \includegraphics[width=\textwidth]{linear_rf_shap_dot_summary.png}
    \caption{Linear nuisance}
\end{subfigure}%
\begin{subfigure}{.5\textwidth}
    \includegraphics[width=\textwidth]{xgb_20k_shap-dot-summary-plot.png}
    \caption{GB nuisance}
\end{subfigure}%
\caption{SHAP summary plot of the DRIV random forest CATE projection model}
\label{fig:shap-dot-2018}
\end{figure}

\begin{figure}[H]
\begin{subfigure}{.5\textwidth}
    \centering
    \includegraphics[width=.95\linewidth]{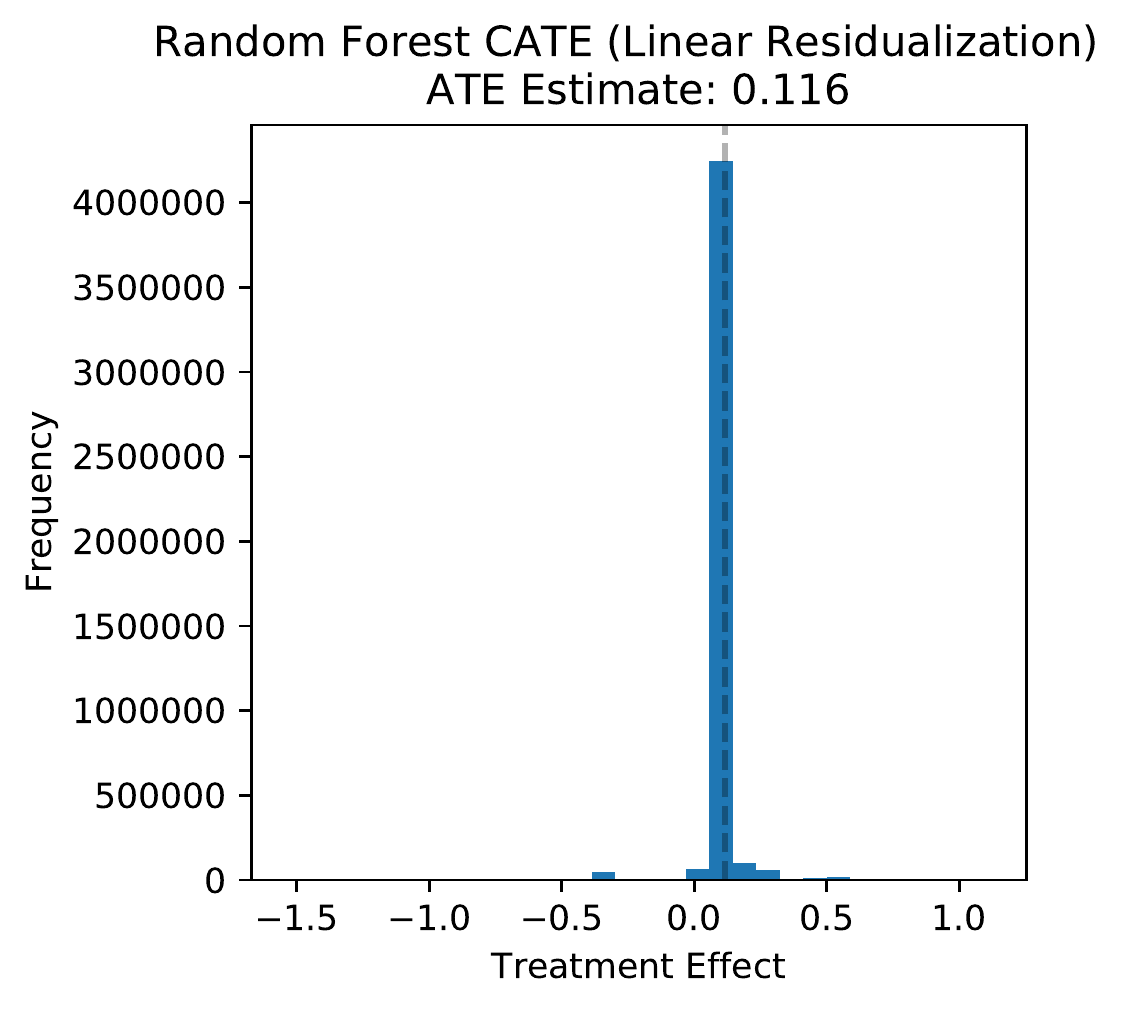}
    \label{fig:rf-cate-w-linear-resid_norm}
\end{subfigure}%
\begin{subfigure}{.5\textwidth}
    \centering
    \includegraphics[width=.9\linewidth]{linear-cate-w-linear-resid_norm.pdf}
    \label{fig:linear-cate-w-linear-resid_norm}
\end{subfigure}
\caption{Distribution of CATE estimates using linear nuisance models.}
\end{figure}

\begin{figure}[H]
\begin{subfigure}{.5\textwidth}
    \includegraphics[width=0.95\textwidth]{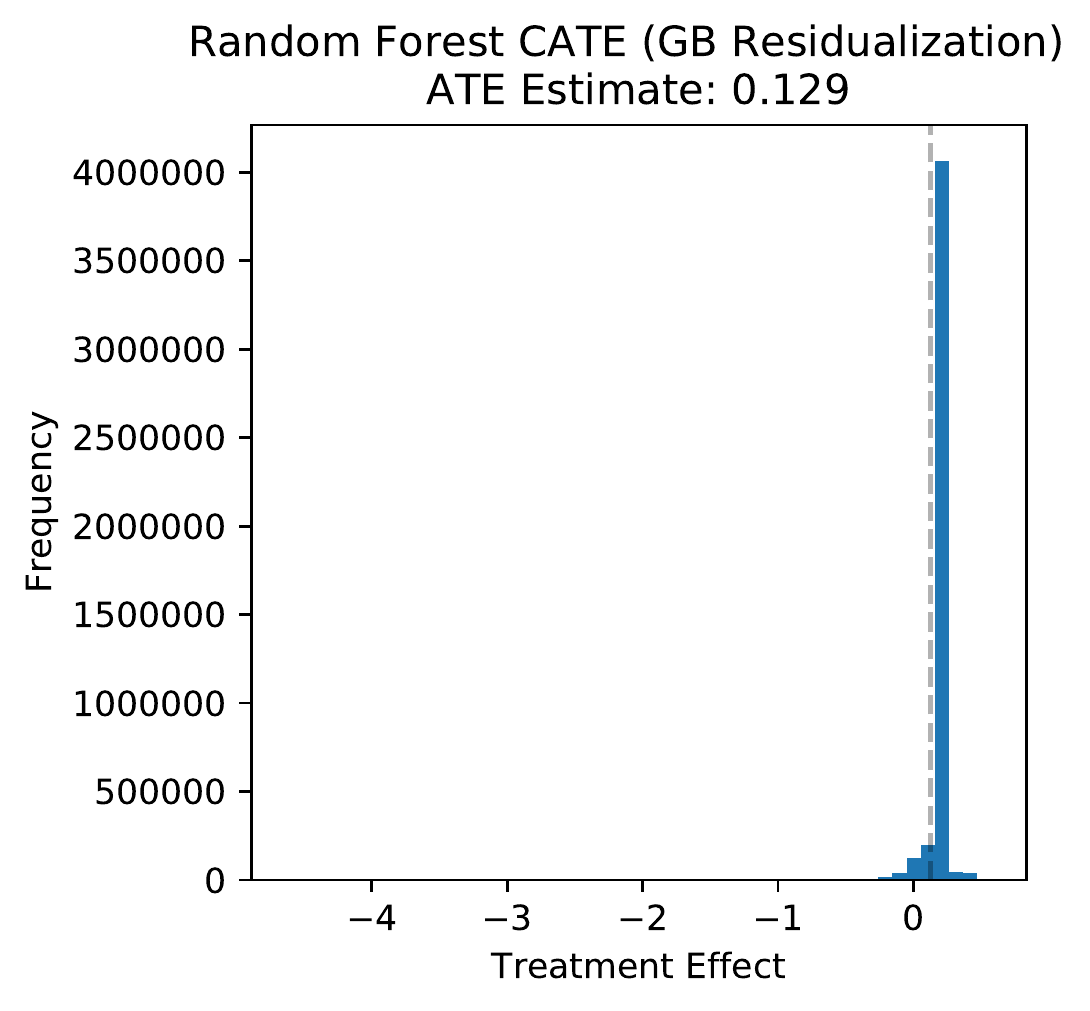}
    \label{fig:rf-cate-w-xgb-resid_norm}
\end{subfigure}%
\begin{subfigure}{.5\textwidth}
    \includegraphics[width=0.9\textwidth]{linear-cate-w-xgb_norm.pdf}
\end{subfigure}
\caption{Distribution of CATE estimates using gradient boosting nuisance models.}
\end{figure}

\section{Semi-Synthetic Data Analysis for TripAdvisor Data}\label{sec:W-semi}
\textbf{TripAdvisor Semi-synthetic Data Results.} In order to validate the correctness of ATE and CATE from DRIV model, we consider a semi-synthetic data generating process that looks similar in structure to TripAdvisor data. The covariates have the same schema but are generated from fixed marginal distributions. The instrument corresponds to a fully randomized recommendation of treatment. And the compliance rates are generated to be similar with the experiment. This probability depends both on the observed feature $X$ and an unobserved confounder that has a direct effect on the outcome. The X covariates and DGP are given by:

\begin{table}[H]
\renewcommand\arraystretch{1.5}
\begin{tabular}{p{0.6\linewidth}|p{0.4\linewidth}} 
\hline
Covariate & Distribution \\
 \hline
days\_visited\_free\_pre, days\_visited\_hs\_pre, days\_visited\_rs\_pre, days\_visited\_exp\_pre, days\_visited\_vrs\_pre, days\_visited\_fs\_pre & $X \sim \text{U}\{0, 28\}$ \\
 \hline
locale\_US & $X \sim \text{Bernoulli}(p=.5)$ \\
 \hline
os\_type & $X \sim \{OSX,Windows,Linux\}$ \\
 \hline
revenue\_pre & $X \sim \text{Lognormal}(\mu=0, \sigma=3)$ \\
 \hline
\end{tabular}
\caption{Data Generation of Covariates X}
\label{table:dgp_x}
\end{table}

\begin{align}
Z \sim \; & \text{Bernoulli}(p=.5) \tag{Instrument}\\
\nu \sim \; & \text{U}[0, 10] \tag{Unobserved confounder}\\
C \sim \; & \text{Bernoulli}(p=0.017\cdot\text{Logistic}(0.1\cdot(X[0] + \nu))) \tag{Compliers when recommended}\\
C0 \sim \; & \text{Bernoulli}(p=0.006) \tag{Non-Compliers when not recommended}\\
T \sim \; & C\cdot Z + C0\cdot (1 - Z) \tag{Treatment}\\
y \sim \; & \theta(X) \cdot (T + 0.1\cdot\nu) + 0.4\cdot X[0] + 2\cdot \text{U}[0, 1] \tag{Outcome}
\end{align}

Moreover, the treatment effect function is predefined here, which depends on the feature "days\_visited\_free\_pre"(X[0]) and "locale\_US"(X[6])
\begin{align}\label{eq:true_te}
\theta(X) = \; & 0.2 + 0.1 \cdot X[0] -2.7 \cdot X[6] \tag{CATE}
\end{align}

\begin{figure}[H]
    \begin{subfigure}[b]{0.3\textwidth}
        \includegraphics[width=\linewidth]{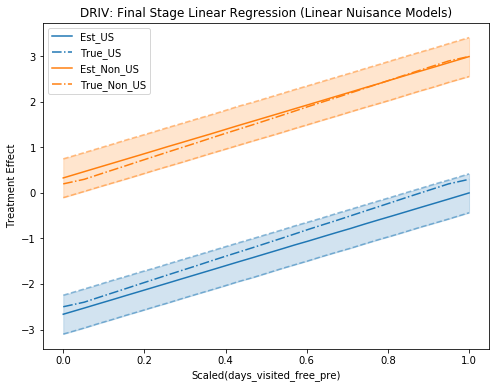} 
        \label{fig:subim1}
    \end{subfigure}
    \begin{subfigure}[b]{0.3\textwidth}
        \includegraphics[width=\linewidth]{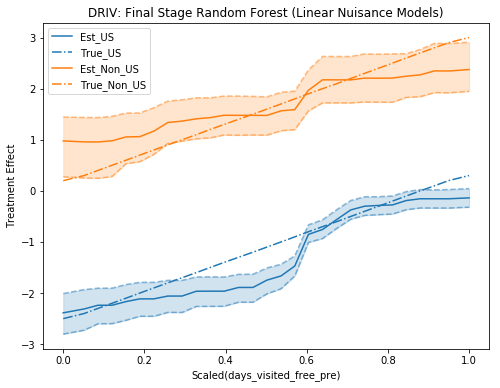} 
        \label{fig:subim2}
    \end{subfigure}
    \begin{subfigure}[b]{0.4\textwidth}
        \includegraphics[width=\linewidth]{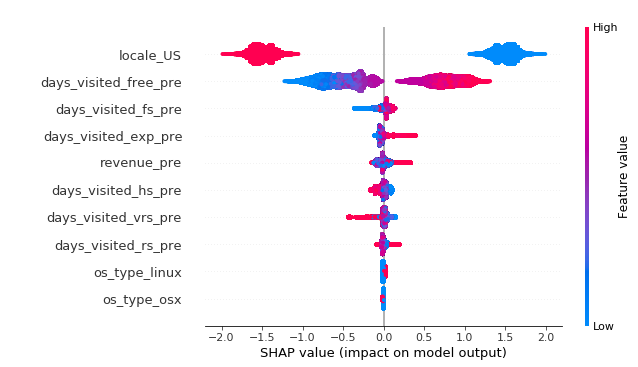}
        \label{fig:subim3}
    \end{subfigure}
\caption{(from left to right) CATE projection on X[0] and X[6] by linear final stage model, CATE projection on X[0] and X[6] by RF final model, SHAP summary of RF CATE projection. (n=4,000,000, coef=0.1, linear nuisance models)}
\label{fig:ta_dgp_linear_1}
\end{figure}

\begin{figure}[H]
    \begin{subfigure}[b]{0.3\textwidth}
        \includegraphics[width=\linewidth]{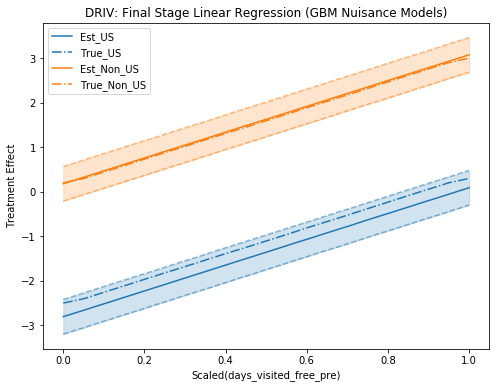} 
        \label{fig:subim1}
    \end{subfigure}
    \begin{subfigure}[b]{0.3\textwidth}
        \includegraphics[width=\linewidth]{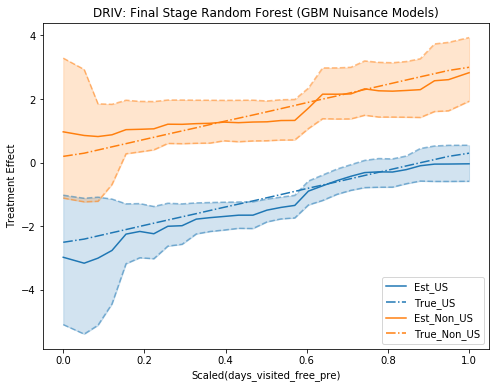} 
        \label{fig:subim2}
    \end{subfigure}
    \begin{subfigure}[b]{0.4\textwidth}
        \includegraphics[width=\linewidth]{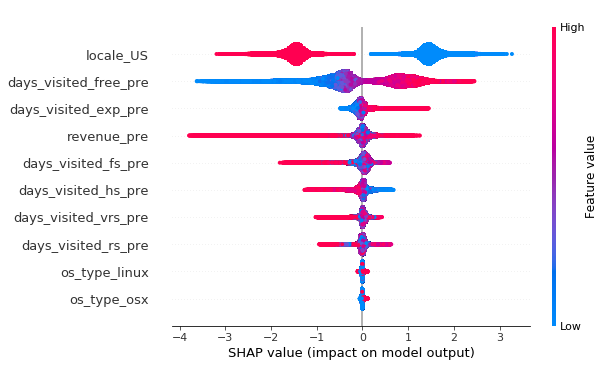}
        \label{fig:subim3}
    \end{subfigure}
\caption{Same plot for n=4,000,000, coef=0.1, GBM nuisance models}
\label{fig:ta_dgp_xgb_1}
\end{figure}

We rerun the same experiments with 4 million samples. In table \ref{table:te_comparison_1}, it shows that both DMLATEIV and DRIV with either linear or GBM nuisance estimators, their ATE CI can recover the true estimate of ATE. Moreover, we validate the CATE via DRIV. In figure \ref{fig:ta_dgp_linear_1} and \ref{fig:ta_dgp_xgb_1}, we can see that DRIV with linear regression as final stage recovers the true coefficient from \ref{eq:true_te}, and the last stage model using random forest also picks the correct factor of heterogeneity as the most important features.  

\begin{table}[H]
\centering
\renewcommand\arraystretch{1.5}
\begin{tabular}{c c c c c} 
 \hline
 Nuisance Models & Method & True ATE & ATE Estimate & 95\% CI \\
 \hline
Linear Models  & DMLATEIV & 0.249 & 0.336 & [0.186, 0.487] \\
Linear Models & DRIV with constant & 0.249 & 0.166 & [-0.025, 0.358] \\
Gradient Boosting Models  & DMLATEIV & 0.249 & 0.342 & [0.191, 0.492 ] \\
Gradient Boosting Models & DRIV with constant & 0.249 & 0.136  & [-0.060, 0.332 ] \\
\end{tabular}
\caption{ATE Estimates for Semi-Synthetic Data (n=4,000,000,coef=0.1)}
\label{table:te_comparison_1}
\end{table}

We also run some other experiments with different sample size $n$ and different level of endogeneity (the coefficient of variable $\nu$) to learn the consistency of ATE for these two models. we can see from the table and figures below that all of their CI covers the true estimate of ATE, but with the increase of $n$ and the decrease of the endogeneity coefficient, the ATE of DMLATEIV is more biased.

\begin{table}[H]
\centering
\renewcommand\arraystretch{1.5}
\begin{tabular}{c c c c c} 
 \hline
 Nuisance Models & Method & True ATE & ATE Estimate & 95\% CI \\
 \hline
Linear Models  & DMLATEIV & 0.249 & 0.349 & [0.230, 0.468] \\
Linear Models & DRIV with constant & 0.249 & 0.197 & [0.044, 0.350] \\
Gradient Boosting Models  & DMLATEIV & 0.249 & 0.354 & [0.235, 0.473] \\
Gradient Boosting Models & DRIV with constant & 0.249 & 0.179  & [0.023, 0.335] \\
\end{tabular}
\caption{ATE Estimates for Semi-Synthetic Data (n=4,000,000,coef=0.01)}
\label{table:te_comparison_2}
\end{table}

\begin{figure}[H]
    \begin{subfigure}[b]{0.3\textwidth}
        \includegraphics[width=\linewidth]{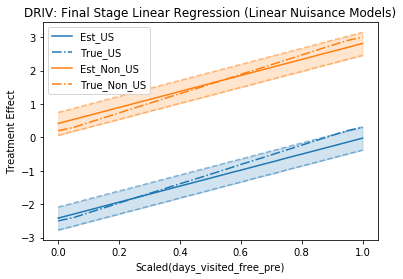} 
        \label{fig:subim1}
    \end{subfigure}
    \begin{subfigure}[b]{0.3\textwidth}
        \includegraphics[width=\linewidth]{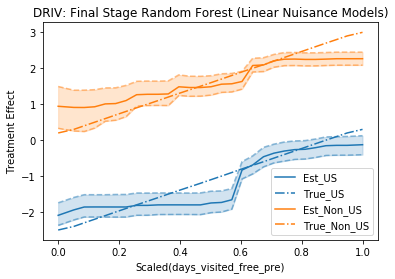} 
        \label{fig:subim2}
    \end{subfigure}
    \begin{subfigure}[b]{0.4\textwidth}
        \includegraphics[width=\linewidth]{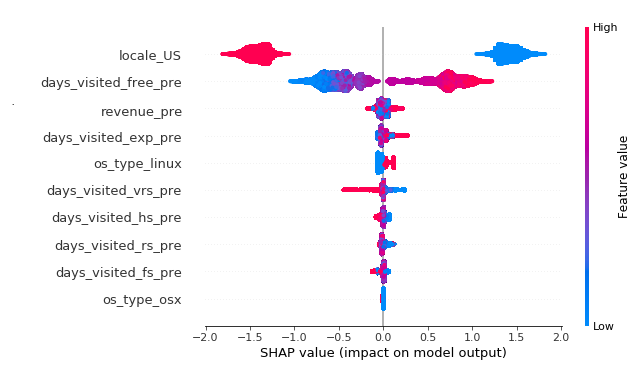}
        \label{fig:subim3}
    \end{subfigure}
\caption{Same plot for n=4,000,000, coef=0.01, Linear nuisance models}
\label{fig:ta_dgp_linear_2}
\end{figure}

\begin{figure}[H]
    \begin{subfigure}[b]{0.3\textwidth}
        \includegraphics[width=\linewidth]{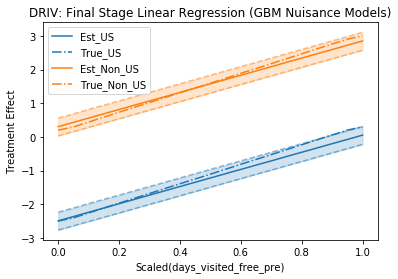} 
        \label{fig:subim1}
    \end{subfigure}
    \begin{subfigure}[b]{0.3\textwidth}
        \includegraphics[width=\linewidth]{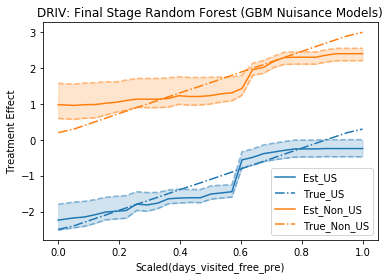} 
        \label{fig:subim2}
    \end{subfigure}
    \begin{subfigure}[b]{0.4\textwidth}
        \includegraphics[width=\linewidth]{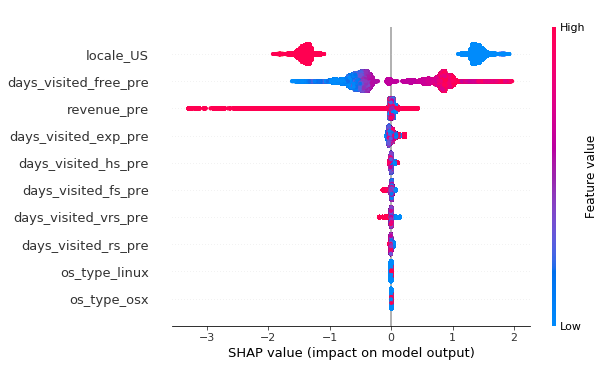}
        \label{fig:subim3}
    \end{subfigure}
\caption{Same plot for n=4,000,000, coef=0.01, GBM nuisance models}
\label{fig:ta_dgp_xgb_2}
\end{figure}

\begin{table}[H]
\centering
\renewcommand\arraystretch{1.5}
\begin{tabular}{c c c c c} 
 \hline
 Nuisance Models & Method & True ATE & ATE Estimate & 95\% CI \\
 \hline
Linear Models  & DMLATEIV & 0.250 & 0.350 & [0.045, 0.655] \\
Linear Models & DRIV with constant & 0.250 & 0.167 & [-0.222, 0.556] \\
Gradient Boosting Models  & DMLATEIV & 0.250 & 0.344 & [0.040, 0.648] \\
Gradient Boosting Models & DRIV with constant & 0.250 & 0.253  & [-0.212, 0.718] \\
\end{tabular}
\caption{ATE Estimates for Semi-Synthetic Data (n=1,000,000,coef=0.1)}
\label{table:te_comparison_3}
\end{table}

\begin{figure}[H]
    \begin{subfigure}[b]{0.3\textwidth}
        \includegraphics[width=\linewidth]{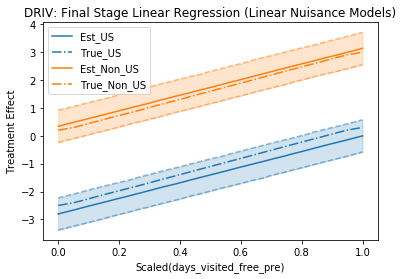} 
        \label{fig:subim1}
    \end{subfigure}
    \begin{subfigure}[b]{0.3\textwidth}
        \includegraphics[width=\linewidth]{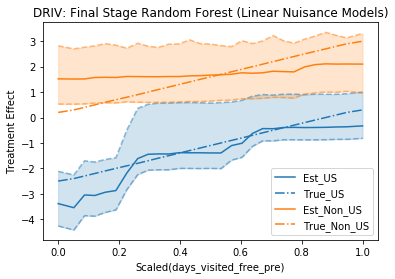} 
        \label{fig:subim2}
    \end{subfigure}
    \begin{subfigure}[b]{0.4\textwidth}
        \includegraphics[width=\linewidth]{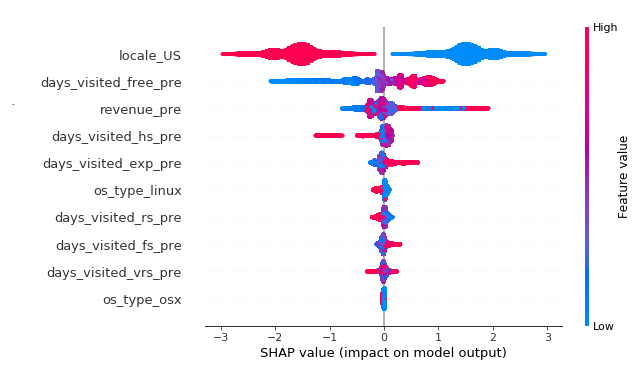}
        \label{fig:subim3}
    \end{subfigure}
\caption{Same plot for n=1,000,000, coef=0.1, Linear nuisance models}
\label{fig:ta_dgp_linear_3}
\end{figure}

\begin{figure}[H]
    \begin{subfigure}[b]{0.3\textwidth}
        \includegraphics[width=\linewidth]{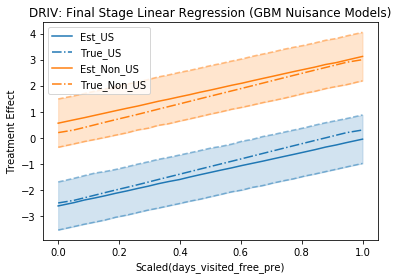} 
        \label{fig:subim1}
    \end{subfigure}
    \begin{subfigure}[b]{0.3\textwidth}
        \includegraphics[width=\linewidth]{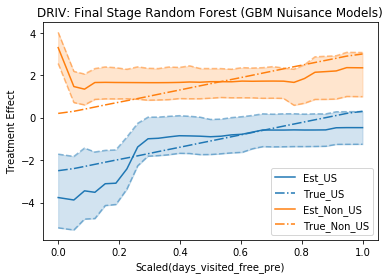} 
        \label{fig:subim2}
    \end{subfigure}
    \begin{subfigure}[b]{0.4\textwidth}
        \includegraphics[width=\linewidth]{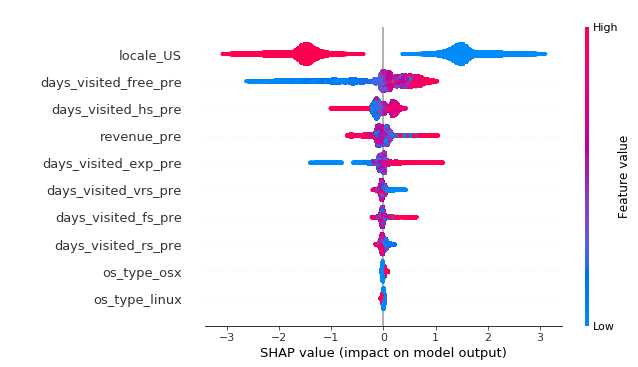}
        \label{fig:subim3}
    \end{subfigure}
\caption{Same plot for n=1,000,000, coef=0.1, GBM nuisance models}
\label{fig:ta_dgp_xgb_3}
\end{figure}

\textbf{Coverage Experiment.} To further validate the consistency of DMLATEIV and DRIV under effect and compliance heterogeneity, we create a slightly different semi-synthetic dataset with stronger instrument and less samples ($n=100,000$) to run 100 times Monte Carlo Simulations. The DGP is given by: 

\begin{align}
Z \sim \; & \text{Bernoulli}(p=.5) \tag{Instrument}\\
\nu \sim \; & \text{U}[0, 10] \tag{Unobserved confounder}\\
C \sim \; & \text{Bernoulli}(p=0.2\cdot\text{Logistic}(0.1\cdot(X[0] + \nu))) \tag{Compliers when recommended}\\
C0 \sim \; & \text{Bernoulli}(p=0.1) \tag{Non-Compliers when not recommended}\\
T \sim \; & C\cdot Z + C0\cdot (1 - Z) \tag{Treatment}\\
y \sim \; & \theta(X) \cdot (T + 0.2\cdot\nu) + 0.1\cdot X[0] + 0.1\cdot \text{U}[0, 1] \tag{Outcome}
\end{align}

Moreover,
\begin{align}
\theta(X) = \; & 0.8 + 0.5 \cdot X[0] -3 \cdot X[7] \tag{CATE}
\end{align}

It turns out that distribution of DMLATEIV ATE has smaller variance but larger bias with 26\% coverage to the true ATE, while DRIV ATE are more converged to the true ATE with 94\% coverage.

\begin{figure}[H]
    \begin{subfigure}[b]{0.33\textwidth}
        \includegraphics[width=\linewidth]{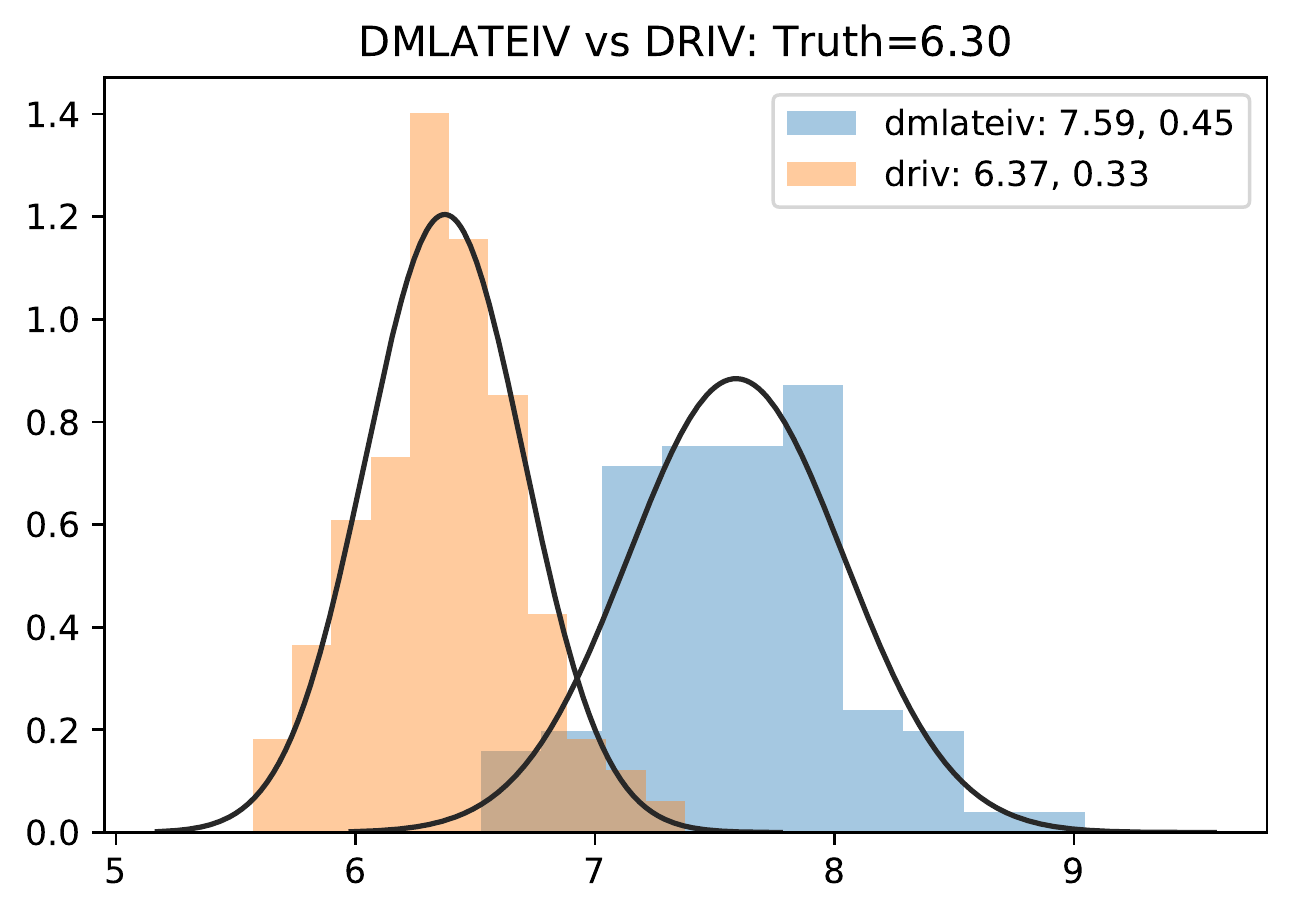} 
        \label{fig:subim1}
    \end{subfigure}
    \begin{subfigure}[b]{0.33\textwidth}
        \includegraphics[width=\linewidth]{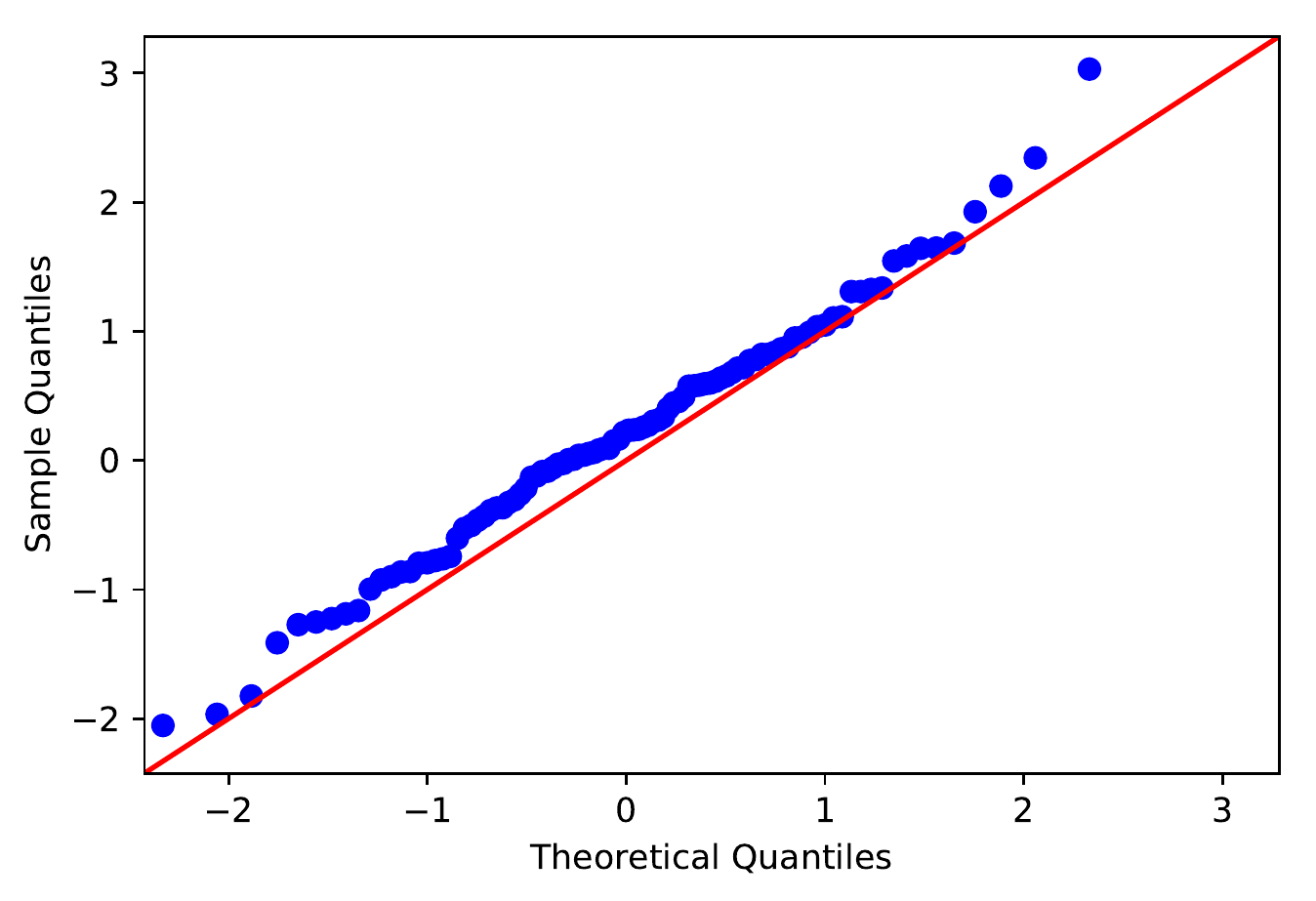} 
        \label{fig:subim2}
    \end{subfigure}
    \begin{subfigure}[b]{0.33\textwidth}
        \includegraphics[width=\linewidth]{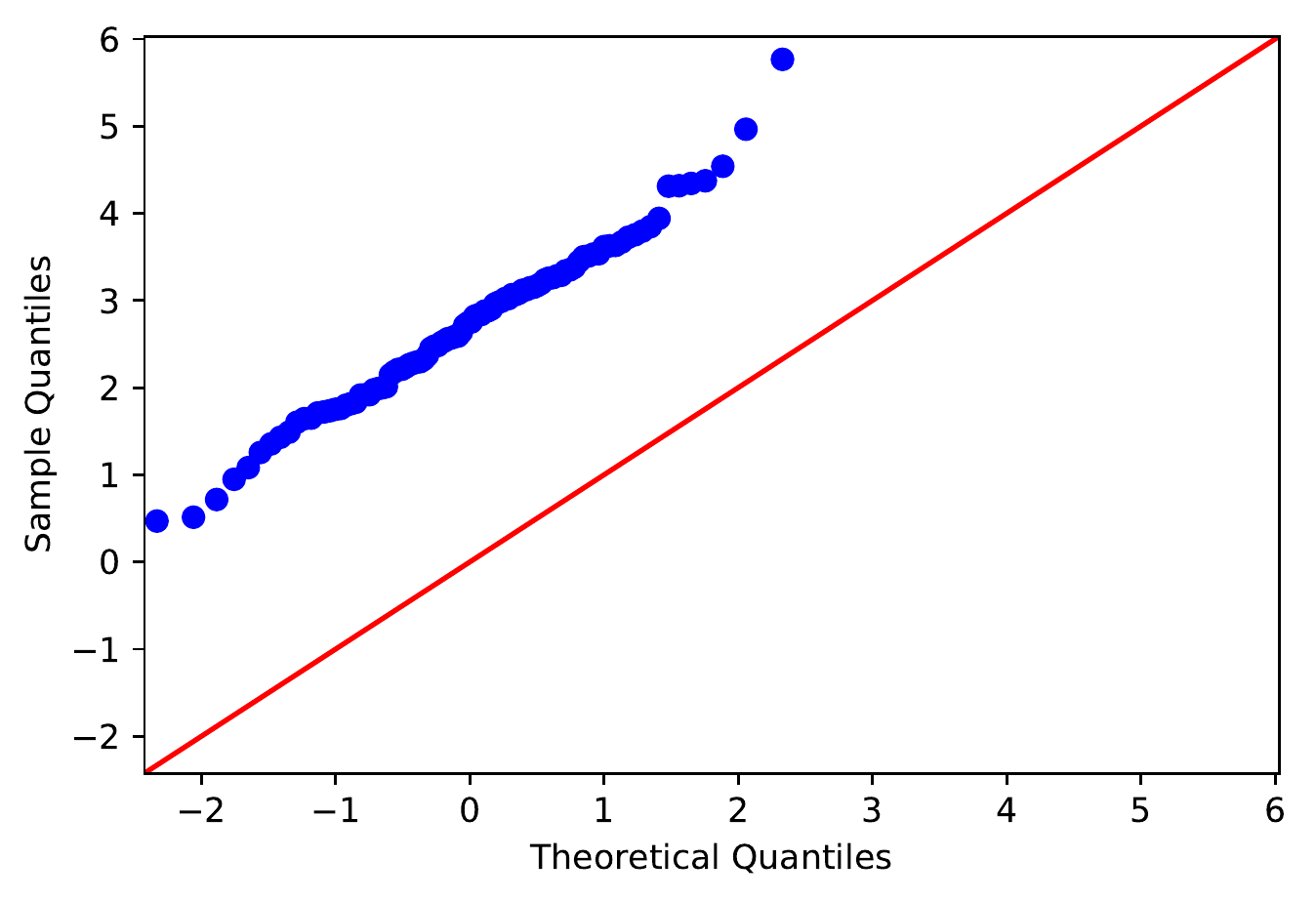}
        \label{fig:subim3}
    \end{subfigure}
\caption{DMLATEIV VS. DRIV ATE Estimates across $100$ Monte Carlo Experiments: (left) distribution of ATEs across experiments, (middle) qq-plot of distribution of DRIV ATE vs normal centered at true estimate, scaled by std of DRIV, (right) qq-plot of distribution of DMLATEIV ATE vs normal centered at true estimate, scaled by std of DMLATEIV.}
\label{fig:cov_exp}
\end{figure}

\section{NLSYM Data Analysis}\label{sec:nlsyc}

\textbf{NLSYM Data Results.} The NLSYM data is comprised of 3,010 entries from men ages 14-24 that were interviewed in 1966 and again in 1976. We use the covariates $X$ selected by Card: mother and father education, family composition at 14, workforce experience, indicators for black, region, southern residence and residence in an SMSA in 1966 and 1976. The outcome of interest $y$ is log wages, the treatment $T$ is the years of schooling, and the instrument $Z$ is an indicator of whether the participant grew up near a 4-year college.

\textbf{Semi-synthetic Data Results.} The NLSYM data is a relatively small dataset and $Z$ could potentially be a weak instrument, which could explain the large confidence intervals in the prior analysis. To disentangle these effects, we create semi-synthetic data from the NLSYM covariates $X$ and instrument $Z$, with generated treatments and outcomes based on known compliance and treatment functions. The data generating process for the semi-synthetic data is given by:
\begin{align}
\nu \sim \; & \text{U}[0, 1] \tag{Unobserved Confounder}\\
C = \; & c_0\cdot X[4], \; c_0 \;(const)\sim \text{U}[0.2, 0.3] \tag{Compliance Level}\\
T = \; & C\cdot Z + g(X) + \nu  \tag{Treatment}\\
y \sim \; & \theta(X) \cdot (T + \nu) + f(X)+\mathcal{N}(0, 0.1) \tag{Outcome}
\end{align}
We create a realistic heterogeneous treatment effect that depends on the mother's education (X[4]) and whether the child was in the care of a single mother at age 14 (X[7], 10\% of subjects):
\begin{align}\label{eq:NLSYM_hte}
\theta(X) = \; & 0.1 + 0.05 \cdot X[4] - 0.1\cdot X[7] \tag{CATE}\\
f(X) = \; & 0.05\cdot X[4], \; g(X) = \; X[4] \tag{Nuissance Functions}
\end{align}

\end{document}